\documentclass[12pt]{article}
\usepackage{amsfonts}
\catcode`\@=11

\input epsf

\@addtoreset{equation}{section}
\def\theequation{\arabic{equation}}
\def\theequation{\thesection\arabic{equation}}

\newcommand{\p}[1]{(\ref{#1})}
\newcommand{\be}{\begin{equation}}
\newcommand{\bea}{\begin{eqnarray}}
\newcommand{\ee}{\end{equation}}
\newcommand{\eea}{\end{eqnarray}}

\def\theequation{\arabic{section}.\arabic{equation}}

\def\pr{\partial}
\def\prd{\partial \cdot}

\def\part{\partial}

\def\dsll{\not {\! \partial}}
\def\psisl{\not {\! \! \psi}}
\def\nablasl{\not {\! \nabla}}

\def\xisl{\not {\! \xi}}
\def\esl{\not {\! \epsilon}}
\def\ssl{\not {\! \cal S}}

\def\@normalsize{\@setsize\normalsize{15pt}\xiipt\@xiipt
\abovedisplayskip 14pt plus3pt minus3pt%
\belowdisplayskip \abovedisplayskip
\abovedisplayshortskip  \z@ plus3pt%
\belowdisplayshortskip  7pt plus3.5pt minus0pt}
\def\small{\@setsize\small{13.6pt}\xipt\@xipt
\abovedisplayskip 13pt plus3pt minus3pt%
\belowdisplayskip \abovedisplayskip
\abovedisplayshortskip  \z@ plus3pt%
\belowdisplayshortskip  7pt plus3.5pt minus0pt
\def\@listi{\parsep 4.5pt plus 2pt minus 1pt
            \itemsep \parsep
            \topsep 9pt plus 3pt minus 3pt}}

\def\underline#1{\relax\ifmmode\@@underline#1\else
        $\@@underline{\hbox{#1}}$\relax\fi}
\@twosidetrue
\relax

\catcode`@=12

\catcode`\@=11

\def\section{\@startsection{section}{1}{\z@}{3.5ex plus 1ex minus
   .2ex}{2.3ex plus .2ex}{\large\bf}}

\def\thesubsection{\arabic{section}.\arabic{subsection}}

\def\ps@headings{\def\@oddfoot{}\def\@evenfoot{}
\def\@oddhead{\hbox{}\hfill
        \makebox[.5\textwidth]{\raggedright\ignorespaces --\thepage{}--
        \hfill }}
\def\@evenhead{\@oddhead}
\def\subsectionmark##1{\markboth{##1}{}} }
\renewcommand{\subsection}[1]{\addtocounter{subsection}{1}
\vspace{2.5mm}\par\noindent {\em \thesubsection . #1}\par
 \vspace{0.5mm} }
\ps@headings

\catcode`\@=12

\relax

\def\figcap{\section*{Figure Captions\markboth
        {FIGURECAPTIONS}{FIGURECAPTIONS}}\list
        {Fig. \arabic{enumi}:\hfill}{\settowidth\labelwidth{Fig. 999:}
        \leftmargin\labelwidth
        \advance\leftmargin\labelsep\usecounter{enumi}}}
 \relax
\def\tablecap{\section*{Table Captions\markboth
        {TABLECAPTIONS}{TABLECAPTIONS}}\list
        {Table \arabic{enumi}:\hfill}{\settowidth\labelwidth{Table 999:}
        \leftmargin\labelwidth
        \advance\leftmargin\labelsep\usecounter{enumi}}}
 \relax
\def\reflist{\section*{References\markboth
        {REFLIST}{REFLIST}}\list
        {[\arabic{enumi}]\hfill}{\settowidth\labelwidth{[999]}
        \leftmargin\labelwidth
        \advance\leftmargin\labelsep\usecounter{enumi}}}
 \relax

\catcode`\@=11

\def\marginnote#1{}
\newcount\hour
\newcount\minute
\newtoks\amorpm
\hour=\time\divide\hour by60
\minute=\time{\multiply\hour by60 \global\advance\minute by-
\hour}
\edef\standardtime{{\ifnum\hour<12 \global\amorpm={am}%
    \else\global\amorpm={pm}\advance\hour by-12 \fi
    \ifnum\hour=0 \hour=12 \fi
    \number\hour:\ifnum\minute<100\fi\number\minute\the\amorpm}}
\edef\militarytime{\number\hour:\ifnum\minute<100\fi\number\minute}
\def\draftlabel#1{{\@bsphack\if@filesw {\let\thepage\relax
  \xdef\@gtempa{\write\@auxout{\string
    \newlabel{#1}{{\@currentlabel}{\thepage}}}}}\@gtempa
    \if@nobreak \ifvmode\nobreak\fi\fi\fi\@esphack}
     \gdef\@eqnlabel{#1}}
\def\@eqnlabel{}
\def\@vacuum{}
\def\draftmarginnote#1{\marginpar{\raggedright\scriptsize\tt#1}}
\def\draft{\oddsidemargin -.5truein
        \def\@oddfoot{\sl preliminary draft \hfil
        \rm\thepage\hfil\sl\today\quad\militarytime}
        \let\@evenfoot\@oddfoot \overfullrule 3pt
        \let\label=\draftlabel
        \let\marginnote=\draftmarginnote

\def\@eqnnum{(\theequation)\rlap{\kern\marginparsep\tt\@eqnlabel}%
\global\let\@eqnlabel\@vacuum}  }
\def\preprint{\twocolumn\sloppy\flushbottom\parindent 1em
        \leftmargini 2em\leftmarginv .5em\leftmarginvi .5em
        \oddsidemargin -.5in    \evensidemargin -.5in
        \columnsep 15mm \footheight 0pt
        \textwidth 250mmin      \topmargin  -.4in
        \headheight 12pt \topskip .4in
        \textheight 175mm
        \footskip 0pt

\def\@oddhead{\thepage\hfil\addtocounter{page}{1}\thepage}
        \let\@evenhead\@oddhead \def\@oddfoot{} \def\@evenfoot{}  }
\def\titlepage{\@restonecolfalse\if@twocolumn\@restonecoltrue\onecolumn
     \else \newpage \fi \thispagestyle{empty}\c@page\z@
        \def\thefootnote{\fnsymbol{footnote}} }
\def\endtitlepage{\if@restonecol\twocolumn \else  \fi
        \def\thefootnote{\arabic{footnote}}
        \setcounter{footnote}{0}}  
\catcode`@=12
\relax

\def\ps@headings{\def\@oddfoot{}\def\@evenfoot{}
\def\@oddhead{\hbox{}\hfill
        \makebox[.5\textwidth]{\raggedright\ignorespaces --\ \thepage{}\ --
        \hfill }}
\def\@evenhead{\@oddhead}
\def\subsectionmark##1{\markboth{##1}{}} }

\ps@headings

\relax

\def\firstpage#1#2#3#4#5#6{
\begin{titlepage}
\nopagebreak
\title{\begin{flushright}
      \vspace*{-0.4in}
        {\normalsize ROM2F-03/31 \ \ \ \ \ }\\[-8mm]
        {\normalsize DFPD/03/TH/43}\\[-8mm]
        {\normalsize hep-th/0311257}\\[4mm]
\end{flushright}
{#3}}
\vskip .1cm
\author{ #4 \\[0.3cm] #5}
\maketitle \vskip -9mm \nopagebreak \vskip .5cm
\begin{abstract} {\noindent #6}
We discuss string spectra in the low-tension limit using the BRST
formalism, with emphasis on the role of triplets of totally
symmetric tensors and spinor-tensors and their generalizations to
cases with mixed symmetry and to (A)dS backgrounds. We also
present simple compensator forms of the field equations for
individual higher-spin gauge fields that display the
{unconstrained} gauge symmetry of a previous non-local
construction and reduce upon partial gauge fixing to the
(Fang-)Fronsdal equations. For Bose fields we also show how a
local Lagrangian formulation with {unconstrained} gauge symmetry
is determined by a previous BRST construction.
\end{abstract}
\vfill
\begin{flushleft}
\today
\end{flushleft}
\end{titlepage}}
\begin{document}
\date{}
\firstpage{3118}{IC/95/34} {\large\bf On higher spins and the
tensionless limit of String Theory} {A. Sagnotti$^{\,a}$ and M.
Tsulaia$^{\,b,}$\footnote{ \sl On leave from Bogoliubov Laboratory
of Theoretical Physics, JINR, 141980 Dubna, RUSSIA and Institute
of  Physics, GAS 380077 Tbilisi, GEORGIA.}} {\small\sl \small\sl
$^{a}$ Dipartimento di Fisica, Universit\`a
di Roma ``Tor Vergata''\\[-5mm]
\small \sl INFN, Sezione di Roma ``Tor Vergata'' \\[-5mm]
\small \sl Via della Ricerca Scientifica 1, \ I-00133 Roma \ ITALY\\[-1mm]
\small\sl$^{b}$ Dipartimento di Fisica, Universit\`a di Padova\\[-5mm]
\small \sl INFN, Sezione di Padova \\[-5mm] \small\sl
Via F. Marzolo 8, 35131 Padova \ ITALY\\[-1mm]
} \vfill\eject
\section{Introduction}

Higher-spin gauge fields are a fascinating topic in Field Theory
that still presents a variety of obscure features and open
problems. To wit, even the basic formulation of their free
dynamics, first proposed long ago by Fronsdal \cite{fronsd} for
Bose fields and by Fang and Fronsdal \cite{fangfronsd} for Fermi
fields, was recently shown to result from a partial gauge fixing
of Maxwell-like or Einstein-like geometric equations
\cite{fs1,fs2} involving the linearized higher-spin curvatures
${\cal R}$ introduced long ago by de Wit and Freedman \cite{dewfr}
(see also \cite{DD}),
\begin{equation}
\frac{1}{\Box^{p}} \ \prd \ {\cal R}^{[p]}{}^{; \alpha_1 \cdots
\alpha_{2p+1}} \  =\  0 \label{oddcurv} \ee for odd spins
$s=2p+1$, and \be \frac{1}{\Box^{p-1}} \ {\cal R}^{[p]}{}^{;
\alpha_1 \cdots \alpha_{2p}} \ =\ 0 \label{evencurv}
\end{equation}
for even spins $s=2p$, where $[p]$ indicates a $p$-fold trace of
the curvatures ${\cal R}^{\mu_1 \ldots \mu_s ; \nu_1 \ldots
\nu_s}$. Only the two familiar (Maxwell and Einstein) cases of
these equations are local, while all others contain
\emph{non-local terms} starting at spin $s=3$. Still, their gauge
fixing to the local Fronsdal form can be attained at the expense
of the trace of the gauge parameter $\Lambda$, denoted by
$\Lambda^{'}$ in the following and constrained to vanish in the
Fronsdal formulation \cite{fs2} along with the double trace of the
gauge field. Strictly speaking, the non-local geometric equations
of \cite{fs1} apply to totally symmetric tensors, a wide and
interesting class of higher-spin gauge fields that does not
exhaust all possibilities in more than four dimensions, but
tensors of mixed symmetry were recently discussed in these terms
in \cite{mixednloc}. Therefore, one can go beyond the Fronsdal
formulation for general tensor gauge fields, eliminating the need
for constraints on the gauge fields themselves or on the gauge
parameters. Still, in order to test the role of the unconstrained
gauge symmetry in the presence of interactions an equivalent
\emph{local} form, obtained combining the basic gauge fields with
suitable \emph{compensators}, appears potentially quite useful. In
\cite{fs1} such a formulation was presented for the relatively
simple case of a spin-3 field, and one of the results of the
present paper is its generalization to symmetric tensors of
arbitrary rank.

Thanks primarily to the work of Vasiliev \cite{fradkvas,vas} (see
also the recent work on higher-dimensional and supersymmetric
extensions by Sezgin and Sundell \cite{ss}), much is now known
about higher-spin interactions, whereas for a long time only
negative results, pointing to the extremely subtle nature of these
systems, have been available. For instance, an early, classic
result in this context was the Aragone-Deser problem, arising in
the presence of gravitational backgrounds with a non-trivial Weyl
tensor \cite{ades}. Since the gauge invariance of the flat-space
Fronsdal Lagrangian rests crucially on the commuting nature of
partial derivatives, the extension to a curved space must face the
potential emergence of terms proportional to the background Weyl
tensor arising from commutators of covariant derivatives, that
would jeopardize the gauge symmetry. Surely enough, such
commutators are present also for lower spins, but they always
combine into Ricci tensors, and for instance supergravity provides
a well-known example of this phenomenon \cite{sugra}. Conformally
flat space times, and in particular the familiar and important
cases of (anti)de Sitter spaces, collectively denoted by (A)dS in
the following, have vanishing Weyl tensors and therefore should
allow the consistent propagation of individual higher-spin fields.
Indeed, the results of \cite{fronsd,fangfronsd} were soon
generalized to (A)dS space-times in \cite{adsfronsd}, but as we
shall see even these cases present some surprising features. In
more general backgrounds, the current understanding is that an
infinite number of such fields in mutual interaction is needed to
define a consistent dynamics.

The work of Vasiliev \cite{vas} (see also \cite{ss}) culminates in
the proposal of consistent interacting higher-spin equations
resulting from the gauging of an infinite-dimensional
generalization of the tangent-space Lorentz algebra that underlies
Einstein's theory in the vielbein formalism. Vasiliev's
construction is also based on the vielbein formalism, whereby the
tangent space algebra is enlarged while only ordinary
diffeomorphisms are left as manifest symmetries. From this
viewpoint, the constraints present in the Fronsdal formulation
only involve algebraic conditions relating tangent-space tensors
to the Minkowski metric, but it is nonetheless interesting to see
whether the extended gauge symmetry of \cite{fs1,fs2} can be
accommodated in a suitable formulation, and gaining some
understanding of this issue was a main motivation for the present
work. In addition, we should stress that the Vasiliev equations,
whereas consistent, are intrinsically non-Lagrangian, since they
lack additional fields needed in an off-shell formulation, and
this is a key open problem in higher-spin dynamics today. Our
results will display simple instances of this type of phenomenon,
since, for instance, the local compensator equations with
unconstrained gauge symmetry we shall meet will also come in two
varieties, a simple and compact non-Lagrangian form and a more
involved off-shell Lagrangian one.

String Theory, to some extent a more familiar system, includes
infinitely many higher-spin massive fields with consistent mutual
interactions, and can provide useful hints on their dynamics, if a
suitable limit where their masses disappear is explored. This is
the low-tension limit, and conversely one can well hope that a
better grasp of higher-spin dynamics could help forward our
current understanding of String Theory, that is mostly based on
its low-spin massless excitations and on their low-energy
interactions.

In this respect, the purpose of this paper is thus twofold. On the
one hand, we describe the ``triplets'', first discussed in 1986 by
A. Bengtsson \cite{ab86} and identified in general in \cite{fs2},
and their generalizations, that make up the full bosonic string
spectrum in the tensionless limit, with special emphasis on the
relatively simple case of fully symmetric tensors, and show how to
extend them to the case of (A)dS backgrounds. These systems
comprise a spin-$s$ field $\varphi$, a spin-($s-1)$ field $C$ and
a spin-$(s-2)$ field $D$, and the corresponding flat-space
equations read
\begin{eqnarray}
&& \Box \; \varphi \ = \ \partial \, C \ , \nonumber \\
&& C \ = \ \partial \cdot \varphi \ - \ \partial \, D \ ,
\nonumber
\\
&& \Box \; D \ = \ \partial \cdot C \ ,
\end{eqnarray}
where, as will often be the case in the following, tensor indices
are left implicit. They propagate a chain of modes of spin $s$,
$s-2$, \ldots, $0$ or $1$ according to whether $s$ is even or odd,
and were also considered in \cite{pt1} as a natural arena for the
BRST technique \cite{BRST}. On the other hand, as we shall see,
their study is rather rewarding, since they provide a direct route
toward the formulation of non-Lagrangian \emph{local} field
equations for higher-spin gauge fields. For instance, in flat
space this local compensator form of the bosonic equations for a
spin-$s$ field $\varphi$ is simply
\begin{eqnarray}
&& {\cal F} \ =\ 3 \, \partial^3 \; \alpha \ , \nonumber \\
&& \varphi^{''} \ = \ 4 \, \partial \cdot \alpha \ + \ \partial \, \alpha^{'} \ ,
\label{compenintr}
\end{eqnarray}
where $\alpha$ a spin-$(s-3)$ compensator. This is
to be compared with the usual Fronsdal equation
\begin{equation}
{\cal F} \ \equiv \ \Box\; \varphi \ - \ \partial \, \partial \cdot \varphi
\ + \ \partial^2 \, \varphi^{'} \ = \ 0 \ ,
\end{equation}
where the gauge field $\varphi$ is subject to the constraint that
its double trace $\varphi^{''}$ vanish identically. We shall
derive these remarkably simple equations for both Bose and Fermi
fields, in flat space and in (A)dS backgrounds, that play a
crucial role in the Vasiliev equations. For Bose fields, we shall
be able to proceed even further, adapting the BRST procedure to
the string in the tensionless limit to extend the compensator
equations (\ref{compenintr}) to a Lagrangian form in flat space.
This result is actually contained in \cite{pt2} where, however, it
was connected to the conventional Fronsdal formulation. Here, on
the other hand, we display its natural link with the unconstrained
gauge symmetry of the non-local geometric equations of
\cite{fs1,fs2}.

Let us stress that the BRST technique, originally conceived as a
tool for quantization in the presence of a gauge symmetry
\cite{BRST}, has proved over the years remarkably powerful also
for formulating classical field theories. This type of
application, initially proposed by Siegel \cite{siegel}, led
promptly to the free String Field Theory constructions of
\cite{szbp}, and shortly thereafter to the extension of the BRST
analysis of \cite{ko} in the presence of open-string interactions
\cite{nnw,witten,sftreview}. More recently, this technique was
widely used in \cite{pt1,pt2,bpt,burdik} to define significant
instances of higher-spin dynamics in flat space and in (A)dS
backgrounds. As we shall see, it has a direct bearing on the
search for extensions of the triplets of \cite{ab86,fs2} and for
the formulation of higher-spin dynamics with an unconstrained
gauge symmetry. These results should be also of some interest in
view of the potential relevance of higher-spin gauge theories
\cite{adscftnew,ss} for the AdS/CFT correspondence \cite{malda} in
the weak gauge-coupling limit, a subject that recently has
received an increasing attention and has also motivated some
authors to reconsider the key properties of low-tension strings
\cite{lowtensionnew}. A related observation is that the BRST
charge of world-sheet reparametrizations embodies a massive
dynamics of the Fronsdal type, some aspects of which are manifest
in the constructions of \cite{BGP,DW,Zinoviev}. However, in this
paper we shall confine our attention to the case of massless
higher spin fields, leaving a detailed BRST analysis of massive
higher spin fields propagating in (conformally) flat backgrounds
for a future study.

Fermi fields also suggest a triplet-like structure, and indeed
some of the excitations present in fermionic strings are described
precisely by the fermionic triplets of symmetric spinor-tensors
proposed in \cite{fs2}. These comprise a spin-$s$ field $\psi$, a
spin-($s-1$) field $\chi$ and a spin-($s-2$) field $\lambda$, and
if the tensor indices are left implicit the corresponding
equations read
\begin{eqnarray}
&& \dsll \psi \ = \ \partial \chi \ , \nonumber \\
&& \dsll \chi \ = \ \partial \cdot \psi \ - \ \partial \lambda  \ , \nonumber \\
&& \dsll \lambda \ =\
\partial \cdot \chi \ .
\end{eqnarray}
Differently from the bosonic triplets, these systems propagate
\emph{all} half-odd integer spin chains up to and including a
given one. After recovering this structure in the NSR string, we
shall be able to deduce from it local compensator equations for
all higher-spin fermions, both in flat space and in an (A)dS
background, although the triplets themselves, for a reason that
will become clear in the BRST analysis presented in Section 5, do
not allow direct Lagrangian (A)dS extensions. Since an off-shell
formulation for higher-spin fermions is being constructed by other
authors \cite{bpnew}, we shall refrain from completing the
relevant steps in this case. All in all, we can well conclude that
the BRST formalism proves once more quite powerful in dealing with
these constrained systems, and provides a straight path toward the
construction of consistent field equations and Lagrangians.

The content of the present paper is as follows. In Section 2 we
discuss how triplets of symmetric tensors emerge from the bosonic
string in the low-tension limit and describe their generalizations
to tensors with mixed symmetry. In Section 3 we present their
extension to (A)dS backgrounds, proceeding in two ways, first by a
direct computation and then adapting the BRST analysis to this
case, since this clarifies the very reason behind the consistency
of the construction. In Section 4 we then turn to local field
equations, in flat space and in (A)dS backgrounds, for individual
higher-spin bosons with the unconstrained gauge symmetry of
\cite{fs1,fs2}, both in a reduced Vasiliev-like form and in a
complete off-shell form motivated by \cite{pt2}. Finally, in
Section 5 we describe the extension of these results to fermions,
recovering the corresponding triplets from the NSR string,
explaining why, rather surprisingly, they do not extend to (A)dS
backgrounds and deriving from them local non-Lagrangian equations
with the unconstrained gauge symmetry of \cite{fs1}.

\section{The bosonic string triplets}

In this Section we describe how the tensionless limit of the free
bosonic string exhibits the triplets of symmetric tensors of
\cite{ab86,fs2}. This does not correspond directly to the behavior
of tensionless strings, where the limit is taken prior to first
quantization, a subject pioneered in \cite{Lindstrom}. We also
display their generalization to the case of tensors with mixed
symmetry, thus completing the description of the open bosonic
string spectrum in the tensionless limit. Actually, although we
shall deal explicitly with the open bosonic string, the closed
bosonic string will be also fully encompassed by our discussion of
generalized triplets in subsection 2.4.

\subsection{The open bosonic oscillators and the Virasoro algebra}

In order to set up our notation, let us begin by recalling some
standard properties of the open bosonic string oscillators, that
in the ``mostly positive'' space-time signature we shall adopt
throughout satisfy the commutation relations
\begin{equation} \label{oscb}
[\alpha^\mu_k, \alpha^\nu_l]\ = \  k \, \delta_{k+l,0} \,
\eta^{\mu \nu} \ .
\end{equation}
The corresponding Virasoro generators
\begin{equation} \label{VB}
L_k \ = \ \frac{1}{2}\, \sum_{l= -\infty}^{+\infty} \,
\alpha^\mu_{k-l}\, \alpha_{\mu l} \ ,
\end{equation}
where $\alpha_0^\mu \, = \, \sqrt{2 \alpha^\prime} \, p^\mu $ and
$p_\mu = - i \partial_\mu$, satisfy the Virasoro algebra
\begin{equation}
[L_k, L_l] \ = \ (k-l)\, L_{k+l} \ + \  \frac{\cal D}{12} \ m\, ( m^2\,
-\, 1 ) \ ,
\end{equation}
where ${\cal D}$ denotes the total space-time dimension.

In this paper we are interested in the tensionless limit, where
the full gauge symmetry of the massive string spectrum is
recovered, and to this end it is convenient to define the reduced
generators \be \ell_0 =  p^2 \ , \qquad \ell_m =  p \cdot \alpha_m
\quad (m \neq 0) \ . \ee They are related by suitable rescalings
to the naive tensionless limit of the Virasoro generators, and
satisfy the simpler algebra \be [\ell_k,\ell_l] \ = \ k\;
\delta_{k+l,\; 0} \ \ell_0 \ , \ee where the central charge has
disappeared.

\subsection{The BRST charge and the tensionless limit}

It is also convenient to introduce the ghost modes $C_k$, of ghost
number $g=1$, and the corresponding anti-ghost modes $B_k$, of
ghost number $g=-1$, with the anti-commutation relations
\begin{equation}
\{C_k, B_{l} \} = \delta_{k+l,\; 0} \ .
\end{equation}
Indeed the BRST operator, first constructed in \cite{ko}, that in
this case is
\begin{equation}
{\cal Q} \ = \ \sum_{-\infty}^{+\infty}
 \left[ C_{-k} \, L_{k} \ - \ \frac{1}{2}
 (k-l) :\, C_{-k}\, C_{-l}\, B_{k+l}:  \right ] \ - \ C_0\ ,
\end{equation}
determines the free string field equation \cite{ko,nnw,witten} \be
{\cal Q} \, | \Phi \rangle \ = \ 0 \ , \label{stringeq} \ee where
for the open bosonic string $| \Phi \rangle$ has ghost number
$g=-1/2$, while the corresponding ghost vacuum satisfies the
conditions
\begin{eqnarray}
&& B_0|0\rangle_{gh}\ = \ 0 \ , \nonumber \\
&& B_k|0\rangle_{gh}\ =\ 0  \quad ( k > 0 ) \ , \nonumber \\
&& C_k |0\rangle_{gh}\ = \ 0  \quad ( k > 0 ) \ ,
\end{eqnarray}
and actually a similar form, with the proper BRST operator,
applies to all types of strings \cite{ms85}.  The nilpotency of
${\cal Q}$ in the critical dimension $({\cal D}=26)$ implies the
existence of an infinite chain of nested gauge symmetries \be
\delta | \Phi \rangle \ = \ {\cal Q} \; | \Lambda \rangle \ ,
\qquad \delta  | \Lambda \rangle \ = \ {\cal Q} \; |
\tilde{\Lambda} \rangle \ , \qquad \ldots \ee that are typical of
systems of forms, and at the same time guarantees the consistency
of eq. (\ref{stringeq}).

Rescaling the ghost variables according to \be c_k \ = \ \sqrt{2\;
\alpha^\prime} \, C_k \ , \qquad b_{k} \ = \ \frac{1}{ \sqrt{2\;
\alpha^\prime}}\ B_k \ee for $k \neq 0$, and as \be c_{0} \ = \
\alpha^\prime \, C_{0}\ , \qquad b_{0} \ =\
\frac{1}{\alpha^\prime}\ B_0 \ee for $k=0$, does not affect their
anti-commutation relations, but allows a non-singular
$\alpha^\prime \rightarrow \infty$ limit that defines the
\emph{identically} nilpotent BRST charge \be \label{Qlimbose} Q \
= \ \sum_{-\infty}^{+\infty} \left[ c_{-k} \, \ell_k \ - \
\frac{k}{2}\, b_0 \, \, c_{-k} \, c_k \right] \ . \ee

We have thus recalled two equivalent manifestations of the
tensionless limit, in the constraint algebra and in the BRST
charge. The latter choice will prove particularly convenient, and
affords interesting generalizations we shall return to repeatedly in the
following sections.

It is convenient to write $Q$ concisely as
\begin{equation} Q \ = \ c_0 \, \ell_0 \ - \ b_0 \, M \ + \
\tilde Q \ , \end{equation}
with
\begin{equation} \label{QM} \tilde Q \ = \ \sum_{k \neq 0} \,
c_{-k} \, \ell_k \quad {\rm and} \quad M\ =\ \frac{1}{2} \,
\sum_{-\infty }^{+ \infty} \, k \ c_{-k} \, c_k \ .
\end{equation}
In a similar fashion, the string field $| \Phi \rangle$ and the
gauge parameter $| \Lambda \rangle$ can be decomposed as
\begin{eqnarray} | \Phi \rangle &=&
\ | \varphi_1 \rangle \ + \ c_0 \, | \varphi_2 \rangle \ , \\
| \Lambda \rangle &=& | \Lambda_1 \rangle \ + \ c_0 \, | \Lambda_2
\rangle \ , \end{eqnarray} and as a result the field equations and
the corresponding gauge transformations become
 \begin{eqnarray}
&& \ell_0 \; |\varphi_1 \rangle \ - \ \tilde Q  \; |\varphi_2
\rangle \ =\ 0 \ , \nonumber
\\ && \tilde Q \; |\varphi_1 \rangle \ - \ M\; |\varphi_2 \rangle \ = \
0 \ , \label{EMNbose} \\
&& \delta |\varphi_1 \rangle \ = \ \tilde Q |\Lambda_1 \rangle \ -
\ M\; |\Lambda_2 \rangle \ ,  \nonumber \\  && \delta |\varphi_2
\rangle \ = \ \ell_0 |\Lambda_1 \rangle \ - \ \tilde Q|\Lambda_2
\rangle \ . \label{NSG2}
\end{eqnarray}

It should be appreciated that these field equations are consistent
and gauge invariant \emph{in any space-time dimension}. This is to
be contrasted with the ordinary equations for the tensile string
where, as is well known, the critical space-time dimension ${\cal
D}=26$ plays a crucial role in allowing a consistent mass
generation.

\subsection{The case of symmetric tensors}

Let us now confine our attention to totally symmetric tensors,
thus working only with the $(\alpha_{-1},\alpha_1)$ oscillator
pair and effectively reducing the constraints to the
$(\ell_{-1},\ell_0,\ell_1)$ triplet. As a result, the string field
$| \Phi \rangle$ and the gauge parameter $| \Lambda \rangle$
involve only the three ghost modes $(c_{-1},c_0,c_1)$ and the
corresponding anti-ghost modes $(b_{-1},b_0,b_1)$, while the ghost
vacuum satisfies the conditions
\be
c_1|0\rangle_{gh}\ = \ 0 \ ,
\qquad b_1|0\rangle_{gh}\ = \ 0 \ , \qquad b_0|0\rangle_{gh}\ =\ 0
\ . \ee
The limiting form of $Q$ then implies that the field
equations describe independent \emph{triplets} $(\varphi,C,D)$ of
symmetric tensors of ranks $(s,s-1,s-2)$, defined via
\begin{eqnarray}
&& |\varphi_1\rangle = \frac{1}{s!}\, \varphi_{\mu_1 \ldots
\mu_s}(x) \alpha_{-1}^{\mu_1} \ldots \alpha_{-1}^{\mu_s}\;
|0\rangle \nonumber \\ && \qquad + \ \frac{1}{(s-2)!}\, D_{\mu_1 \ldots
\mu_{s-2}}(x) \alpha_{-1}^{\mu_1}
\ldots \alpha_{-1}^{ \mu_{s-2}} \, c_{-1} \, b_{-1} \; |0\rangle \ , \nonumber \\
&& |\varphi_2 \rangle \ = \ \frac{-i}{(s-1)!}\, C_{\mu_1 \ldots
\mu_{s-1}}(x) \alpha_{-1}^{\mu_1}  \ldots \alpha_{-1}^{ \mu_{s-1}}
\, b_{-1} \; |0\rangle \ ,
\end{eqnarray}
while the corresponding gauge transformation parameters $|\Lambda
\rangle$,
\begin{equation}
|\Lambda\rangle \ =\ \frac{i}{(s-1)!}\,
\Lambda_{\mu_1\mu_2...\mu_{s-1}}(x) \, \alpha_{-1}^{\mu_1} \ldots
\alpha_{-1}^{\mu_{s-1}} b_{-1} \, |0\rangle \ ,
\end{equation}
describe symmetric tensors of rank $(s-1)$.

In dealing with these systems of symmetric tensors, it is
convenient to resort to the compact notation of \cite{fs1,fs2},
thus omitting all indices carried by the totally symmetric triplet
fields, by the Minkowski metric tensor and by space-time
derivatives. One can then proceed rather simply, but for a few
seemingly unfamiliar combinatoric rules \cite{fs1,fs2}, that we
collect for later use,
\begin{eqnarray}
&& \left( \partial^{\; p} \ \varphi  \right)^{\; \prime} \ = \ \Box \
\partial^{\; p-2} \
\varphi \ + \ 2 \, \partial^{\; p-1} \  \partial \cdot \varphi \ + \
\partial^{\; p} \
\varphi {\;'} \ , \\
&& \partial^{\; p} \ \partial^{\; q} \ = \ \left( {p+q} \atop p
\right) \
\partial^{\; p+q} \ ,
\\
&& \partial \cdot  \left( \partial^{\; p} \ \varphi \right) \ = \
\Box \
\partial^{\; p-1} \ \varphi \ + \
\partial^{\; p} \ \partial \cdot \varphi \ ,  \\
&& \partial \cdot  \eta^{\;k} \ = \ \partial \, \eta^{\;k-1} \ , \\
&& \left( \eta^k \, T_{(s)} \,  \right)^\prime \ = \ k \, \left[
\, {\cal D} \, + \, 2(s+k-1) \,  \right]\, \eta^{k-1} \, T_{(s)} \ + \
\eta^k \,  T_{(s)}^\prime \ , \label{etak}
\end{eqnarray}
where ${\cal D}$ denotes the total space-time dimension and $T_{(s)}$ is
a generic symmetric rank-$s$ tensor.

Expanding (\ref{EMNbose}) and (\ref{NSG2}) then leads to the
\emph{triplet} equations of \cite{fs2}
\begin{eqnarray}
&&\Box \; \varphi \ = \ \partial \, C \ , \nonumber \\
&& \partial \cdot \varphi \ - \ \partial \, D \ = \ C \ , \nonumber \\
&& \Box \; D \ = \ \partial \cdot C \ , \label{flattriplet}
\end{eqnarray}
and to the corresponding gauge transformations
\begin{eqnarray}
&& \delta \varphi \ = \ \partial \, \Lambda \ , \nonumber \\
&& \delta  C \ = \ \Box \; \Lambda \ , \nonumber \\
&& \delta D \ = \ \partial \cdot \Lambda \ .
\label{flattripletgauge}
\end{eqnarray}
Let us stress that here $\Lambda$ is an \emph{unconstrained}
parameter, to be contrasted with the \emph{traceless} gauge
parameter of the Fronsdal formulation of higher-spin gauge fields
\cite{fronsd}. Interestingly, this type of structure was first
exhibited long ago in the tensionless limit of the open bosonic
string, for the first two massive levels, by A. Bengtsson
\cite{ab86}, in an equivalent form without the field $C$. As
discussed in \cite{ab86,fs2}, these equations propagate modes of
spin $s$, $s-2$, $...$, down to zero or one according to whether
$s$ is even or odd.  This makes up a total of $\left( {{\cal D} +
s - 3}\atop s \right)$ degrees of freedom if ${\cal D} > 4$, or
simply $(s+1)$ degrees of freedom if ${\cal D} = 4$. Nevertheless,
as we shall see, these systems have a lot to teach us about
irreducible higher-spin propagation. They were also considered in
\cite{pt1} as a particularly simple application of the BRST
technique to describe massive fields via dimensional reduction.

It is interesting to note that the combinatorial identity
\begin{equation}
\left( {{\cal D} + s - 2}\atop s \right) \ = \ \sum_{k=0}^s \
\left( {{\cal D} + k - 3}\atop k \right)
\end{equation}
suggests a mechanism of mass generation whereby a triplet gains
mass swallowing a chain of other triplets of lower maximum spins.
However, while such a nice and simple mechanism indeed applies to
the massive Kaluza-Klein modes originating from a ${\cal D}+1 \to
{\cal D}$ reduction, it cannot be held directly responsible for
the mass generation in String Theory, where the mechanism takes
place also within a single triplet, so that in fact the triplet
structure is well hidden in tensile string spectra.

These field equations follow from the Lagrangian
\begin{equation} \label {LBRST}
{\cal L} \ = \ \langle \Phi |\, Q \, |\Phi
\rangle \ ,
\end{equation}
that in component notation reads
\begin{eqnarray}
{\cal L} &=& - \, \frac{1}{2}\ (\partial_\mu \varphi)^2 \ + \ s\,
\partial \cdot \varphi \, C \ + \ s(s-1)\, \partial \cdot C \, D \ \nonumber \\
&+&  \ \frac{s(s-1)}{2} \ (\partial_\mu D)^2 \ - \ \frac{s}{2} \,
C^{\; 2} \ , \label{LtripletB}
\end{eqnarray}
where the $D$ field, whose modes disappear on the mass shell, has
a peculiar negative kinetic term. Alternatively, one can eliminate
the auxiliary field $C$, thus obtaining an equivalent formulation
in terms of \emph{pairs} $(\varphi,D)$ of symmetric tensors, more
in the spirit of \cite{ab86}. In terms of the Fronsdal kinetic
operators \be {\cal F} \ = \ \Box \; \varphi \ - \
\partial \; \partial \cdot \varphi \ + \ \partial^{\; 2} \; \varphi^{'}
\ , \label{fronsdalflat} \ee that satisfy the Bianchi identities
\be
\partial \cdot {\cal F} \ - \ \frac{1}{2} \, \partial \, {\cal
F}^{'} \ = \ - \ \frac{3}{2} \, \partial^{\;3} \, \varphi^{\; ''}
\ , \label{bianchiflat} \ee the field equations then become
\begin{eqnarray}
&&{\cal F} \ = \ \partial^2 \, \left( \varphi^{'} - 2D\right) \ , \nonumber \\
 && \Box \; D \ = \ \frac{1}{2} \, \partial \cdot \partial \cdot
 \varphi \ - \ \frac{1}{2} \, \partial \ \partial \cdot D \ ,
 \label{tripletnoB}
\end{eqnarray}
and follow from the Lagrangians
\begin{eqnarray}
{\cal L} &=& - \, \frac{1}{2}\ (\partial_\mu \varphi)^2 \ + \
\frac{s}{2} \,
(\partial \cdot \varphi)^2 \ + \ s(s-1)\, \partial \cdot \partial \cdot \varphi \, D \ \nonumber \\
&+& \ s(s-1) \ (\partial_\mu D)^2 \ + \ \frac{s(s-1)(s-2)}{2} \,
(\partial \cdot D)^2  \ . \label{LtripletnoB}
\end{eqnarray}

\subsection{Generalized bosonic triplets of mixed symmetry}

It is actually not difficult to account for more general tensors
of mixed symmetry resulting from the interplay of $r$ types of
string oscillators. \footnote{An early BRST treatment of these
systems may be found in \cite{hennteit}. We are grateful to G.
Bonelli for calling this reference to our attention.} The totally
symmetric rank-$s$ field $\varphi$ is then replaced by more
general gauge fields $\varphi$ with $r$ sets of ${n_1, \ldots ,
n_r}$ totally symmetric indices, such that $\sum_{k=1}^r \; n_k =
s$, and the resulting system will describe a total of
\begin{equation}
\prod_{k=1}^r \ \left( {{{\cal D} + n_k - 3}\atop n_k} \right)
\end{equation}
degrees of freedom. The natural guess would then be that the
single auxiliary $C$ field be replaced by $r$ auxiliary fields
$C^i$ $(i=1,\ldots,r)$, and finally that the single $D$ field be
replaced by $r^2$ additional fields $D_i^j$ $(i,j=1,\ldots,r)$.

The resulting gauge transformations should thus be
\begin{eqnarray}
&& \delta \, \varphi \ = \ \sum_{i=1}^r \, \partial^i \, \Lambda^i
\ ,
\nonumber \\
&& \delta \, C^i \ = \ \Box \, \Lambda^i \ , \nonumber \\
&& \delta \, D^{ij} \ = \ \,   \partial^i \cdot \Lambda^j \ ,
\label{gentrip}
\end{eqnarray}
where $\partial^i$ denotes a derivative with respect to an index
of the $i$-th set, so that the corresponding field equations
\begin{eqnarray}
&& \Box \, \varphi \ = \ \sum_{i=1}^r \partial^i \; C^i \ , \nonumber \\
&& \partial^i \cdot \varphi \ - \ \sum_{j=1}^r \, \partial^j \,
D^{ij} \ = \ C^i \ , \nonumber \\
&& \Box \, D^{ij} \ = \  \ \partial^i \cdot C^j \ ,
\label{gentriplet}
\end{eqnarray}
would be the natural generalization of eq.~(\ref{flattriplet}).

The proper description of this system, however, requires a
constraint,
\begin{equation}
\partial^k \cdot D^{ij} = \partial^i \cdot D^{kj} \ ,
\label{gentrconst}
\end{equation}
whose emergence can be anticipated since the two apparently
distinct expressions transform identically under the gauge
transformations (\ref{gentrip}). This constraint, instrumental in
attaining the elimination of all unwanted field components, has an
important consequence: in general (\ref{gentriplet}) and
(\ref{gentrconst}) \emph{are not Lagrangian equations}, since one
is missing at least the Lagrange multipliers needed to enforce
(\ref{gentrconst}). This is the first instance of a phenomenon
that we shall meet again in the following, since indeed only
weaker conditions follow from the integrability of the second of
(\ref{gentriplet}).

For instance, for a field $\varphi_{\mu\nu,\rho\sigma}$, symmetric
only under the interchange of the indices within the two sets,
eqs. (\ref{gentriplet}) become
\begin{eqnarray}
&& \Box \; \varphi_{\mu\nu,\rho\sigma} \ = \ \partial_\mu \,
C^1_{\nu,\rho\sigma} + \partial_\nu \, C^1_{\mu,\rho\sigma} +
\partial_\rho \,
C^2_{\mu\nu,\sigma} + \partial_\sigma \, C^2_{\mu\nu,\rho}
\ , \nonumber \\
&& C^1_{\nu,\rho\sigma} \ = \ \partial^\mu \;
\varphi_{\mu\nu,\rho\sigma} \, - \,
\partial_\nu D^{11}_{\rho\sigma} \, - \, \partial_\rho D^{12}_{\nu,\sigma}
\, - \, \partial_\sigma D^{12}_{\nu,\rho}  \ , \nonumber \\
&& C^2_{\mu\nu,\sigma} \ = \ \partial^\rho \;
\varphi_{\mu\nu,\rho\sigma} \, - \,
\partial_\sigma D^{22}_{\mu\nu} \, - \, \partial_\mu D^{21}_{\nu,\sigma}
\, - \, \partial_\nu D^{21}_{\mu,\sigma}  \ , \nonumber \\
&& \Box \; D^{11}_{\rho\sigma} \ = \ \partial^\nu\; C^1_{\nu,\rho\sigma} \ , \nonumber \\
&& \Box \; D^{12}_{\nu,\sigma} \ = \ \partial^\mu \; C^2_{\mu\nu,\sigma} \ , \nonumber \\
&& \Box \; D^{21}_{\nu,\sigma} \ = \ \partial^\rho\; C^1_{\nu,\rho\sigma} \ , \nonumber \\
&& \Box \; D^{22}_{\mu\nu} \ = \ \partial^\sigma \;
C^2_{\mu\nu,\sigma} \ ,
\end{eqnarray}
where each ',' separates two different groups of totally symmetric
space-time indices. The second and third of these then lead to the
integrability constraint
\begin{equation}
\partial_\nu \partial^\rho \; D^{11}_{\rho\sigma} \, + \, \partial^\rho \partial_\sigma \;
D^{12}_{\nu,\rho} \ = \ \partial^\mu \partial_\sigma \;
D^{22}_{\mu\nu} \, + \, \partial^\mu \partial_\nu \;
D^{21}_{\mu,\sigma} \ , \label{integrab}
\end{equation}
weaker than eq. (\ref{gentrconst}), that in this case would lead
to the two conditions
\begin{eqnarray}
&& \partial^\mu \; D^{21}_{\mu,\sigma} \ = \ \partial^\rho \;
D^{11}_{\rho\sigma} \ , \nonumber \\
&& \partial^\mu \; D^{22}_{\mu\nu} \ = \ \partial^\rho \;
D^{12}_{\nu,\rho} \ , \label{naiveconstr}
\end{eqnarray}
that clearly imply (\ref{integrab}).

The BRST technique leads nicely to a solution of the problem and
to an off-shell formulation for these generalized triplets, albeit
in terms of a wider set of fields. To this end, one has to resort
to a family of $\alpha_{-i}^\mu$ oscillators ($i=1,\ldots,r)$,
that are needed to build tensors of this general type, and these
bring about corresponding (anti)ghosts ($b_{\pm i}$)$c_{\pm i}$.
The result is a wider, but still \emph{finite}, set of fields that
generalize the naive $\varphi$, $C_i$ and $D_{ij}$, collectively
written as
\begin{equation}
|\Phi \rangle \ = \ \frac{c_{-i_1} \, ... \, c_{-i_l}\, b_{-j_1}\,
.. \, b_{-j_l}}{{(l!)}^2}\, |D_{i_1...i_l}^{j_1...j_l} \rangle \
+ \ \frac
 {c_0 \, c_{-i_i}\,
.. \, c_{-i_{l-1}}\, b_{-j_1}\, ... \, b_{-j_l}}{(l-1)! \, l!}\,
|C_{i_1...i_{l-1}}^{j_1...j_l} \rangle \ ,
\end{equation}
where each $C$ and $D$ ``ket'' depends on the bosonic oscillators
$\alpha_{-i}^\mu$, and where the original $\varphi$ field is
described by the $| D \rangle$ field carrying no (anti)ghost
indices, while the original $C_i$ are described by the $| C
\rangle$ fields with the lowest number of indices, that in their
case is indeed a single anti-ghost index due to the $c_0$ factor.
The individual terms contain variable numbers of $\alpha_{-i}$
oscillators, as required by the structure of the field $\varphi$
one is trying to describe, so that each index $i_p$ (or $j_p$)
carried by the $C$ or $D$ fields reduces by one unit the
corresponding number of $\alpha_{-i_p}$ (or $\alpha_{-j_p}$)
oscillators.

 The corresponding gauge parameters can be collectively written
\begin{eqnarray}
|\Lambda^{(1)} \rangle &=& \frac
 {c_{-i_1} ... c_{-i_l}\; b_{-j_1} ...
b_{-j_{l+1}}}{l!(l+1)!}\; |{\Lambda^{1(1)\;
j_1...j_{l+1}}_{i_1,...i_l}} \rangle \nonumber \\ &+&
 \frac{c_0 \; c_{-i_1}
.. c_{-i_{l-1}}\; b_{-j_1} ... b_{-j_{l+1}}}{(l-1)!(l+1)!}\;
|{\Lambda^{2(1)\; j_1...j_{l+1}}_{i_1...i_{l-1}}}
 \rangle \ ,
\end{eqnarray}
to distinguish them from the ``gauge-for-gauge'' parameters
$|\Lambda^{(p)} \rangle$ (with $p>1$), that are now present. The
resulting field equations
\begin{eqnarray}
&& \ell_0\; |D_{i_1...i_l}^{j_1...j_l} \rangle \, +\,  {(-1)}^l
\ell_{i_l}\; |C_{i_1...i_{l-1}}^{j_1...j_l} \rangle \, -\,
{(-1)}^l
\ell_{-j}\; |C_{i_1...i_l}^{j,j_1...j_l} \rangle \ = \ 0 \ , \nonumber \\
&& \ell_{i_l}\; |D_{i_1,...i_{l-1}}^{j_1,..j_{l-1}} \rangle \, -\,
\ell_{-j}\; |D_{i_1...i_{l}}^{j,j_1...j_{l-1}} \rangle \, +\,
{(-1)}^l\; |C_{i_1...i_{l-1}}^{i_l j_1 ...j_{l-1}} \rangle \ = \ 0
\ ,
\end{eqnarray}
are thus invariant under the gauge transformations
\begin{eqnarray}
\!\!\!\!\!\! && \delta \; |D_{i_1...i_l}^{j_1...j_l} \rangle \ =\
- \, {(-1)}^l\ell_{i_l}\; |\Lambda^{1(1)\; j_1.
.j_{l}}_{i_1...i_{l-1}} \rangle \nonumber + {(-1)}^l
 \ell_{-j}\; |\Lambda^{1(1)\; j j_1...j_{l}}_{i_1...i_{l}}
\rangle \, - \, |\Lambda^{2(1)\; i_l j_1 ...j_{l}}_{i_1...i_{l-1}}
\rangle \ , \\
\!\!\!\!\!\! && \delta \; |C_{i_1...i_{l-1}}^{j_1...j_l} \rangle =
\ell_0\; |\Lambda^{1(1)\; j_1...j_l}_{i_1...i_{l-1}} \rangle -
{(-1)}^l \ell_{i_{l-1}}\; |\Lambda^{2(1)\;
j_1...j_l}_{i_1...i_{l-2}} \rangle + {(-1)}^l \ell_{-j}\;
|\Lambda^{(1)\; j j_1...j_l}_{i_1...i_{l-1}} \rangle \ ,
\end{eqnarray}
that, in their turn, are invariant under the chain of
``gauge-for-gauge'' transformations
\begin{eqnarray}
 \delta \; |\Lambda^{1(k)\; j_1...j_{l+k}}_{i_1...i_l} \rangle &=& -
{(-1)}^l\ell_{i_l}\; |\Lambda^{1(k+1)\;
j_1...j_{l+k}}_{i_1...i_{l-1}} \rangle  \nonumber \\ &+& {(-1)}^l
 \ell_{-j}\; |\Lambda^{1(k+1)\; j j_1..j_{l+k}}_{i_1...i_{l}}
\rangle \, - \, |\Lambda^{2(k+1)\; i_l
j_1...j_{l+k}}_{i_1...i_{l-1}}
\rangle \ , \nonumber \\
\delta \; |\Lambda^{2(k)\; j_1...j_{l+k}}_{i_1...i_{l-1}} \rangle
&=& \ell_0 \; |\Lambda^{1(k+1)\; j_1...j_{l+k}}_{i_1...i_{l-1}}
\rangle \nonumber \\ &-& {(-1)}^l \ell_{i_{l-1}}\;
|\Lambda^{2(k+1)\; j_1...j_{l+k}}_{i_1...i_{l-2}} \rangle +
{(-1)}^l \ell_{-j}\; |\Lambda^{2(k+1)\; j
j_1...j_{l+k}}_{i_1...i_{l-1}} \rangle \ ,
\end{eqnarray}
and so on.

Partial gauge fixing of this system reduces it to
(\ref{gentriplet}), but the BRST technique provides directly an
off-shell description, and leads to the gauge-invariant
Lagrangians
\begin{eqnarray}
{\cal L} &=& - \ \sum_l \ \left[ \frac{(-1)^l}{{(l!)}^2} \langle
D_{j_1...j_l}^{i_1...i_l}|\; \ell_0\; |D_{i_1...i_l}^{j_1...j_l}
\rangle \ + \ \frac{2}{(l-1)!\; l!}  \; \langle
C_{i_1...i_{l-1}}^{j_1...j_l}| \;
\ell_{-i_l} \; | D_{j_1...j_l}^{i_1...i_l} \rangle  \right. \nonumber \\
&+& \left. \frac{ 2\; {(-1)^l}}{{(l!)}^2} \langle
C_{i_1...i_{l}}^{j_i...j_{l+1}}|\; \ell_{j_{l+1}} \;
|D_{j_1...j_l}^{i_1...i_l} \rangle \ -\
\frac{{(-1)}^l}{{((l-1)!)}^2} \langle
C_{j_1...j_{l-1}}^{i_1...i_{l-1}j_l}\; |\;
 C_{i_1...i_{l-1}}^{j_1...j_l} \rangle \right] \ .
\end{eqnarray}

For the tensor $\varphi_{\mu\nu,\rho\sigma}$ considered above, the
BRST analysis introduces the previous fields
$C^1_{\nu,\rho\sigma}$, $C^2_{\mu\nu,\rho}$,$D^{11}_{\rho\sigma}$,
$D^{12}_{\nu,\sigma}$, $D^{21}_{\mu,\rho}$ and $D^{22}_{\mu\nu}$,
together with the additional ones, $C^{112}_\rho$, $C^{122}_\mu$
and $D^{1212}$, so that the resulting equations include
(\ref{gentriplet}), that are not modified, together with
additional ones,
\begin{eqnarray}
&&  \partial^\rho\; D^{11}_{\rho\sigma} \, - \, \partial^\nu\;
D^{21}_{\nu\sigma} \ = \ C^{112}_\sigma +
\partial_\sigma\; D^{1212} \ , \nonumber\\
&& \partial^\rho\; D^{12}_{\nu\rho} \, - \, \partial^\mu\;
D^{22}_{\mu\nu} \ = \ C^{122}_\nu \ - \ \partial^\nu\; D^{1212} \
, \nonumber \\
&& \Box \; D^{1212} \ = \ \partial^\rho \; C^{112}_\rho \ - \
\partial^\mu \; C^{122}_\mu  \ , \label{BRSTgentr}
\end{eqnarray}
for the new fields $C^{112}_\rho$, $C^{122}_\mu$ and $D^{1212}$.
Eqs. (\ref{gentriplet}) and (\ref{BRSTgentr}) are now invariant
under the modified gauge transformations
\begin{eqnarray}
&& \delta \; \varphi_{\mu\nu,\rho\sigma} \ = \ \partial_\mu \;
\Lambda^1_{\nu,\rho\sigma} \ + \ \partial_\nu \;
\Lambda^1_{\mu,\rho\sigma} \ + \ \partial_\rho \;
\Lambda^2_{\mu\nu,\sigma} \ + \ \partial_\sigma \;
\Lambda^2_{\mu\nu,\rho}\nonumber \\
&& \delta \; C^1_{\mu,\rho\sigma} \ = \Box \Lambda^1_{\mu,\rho\sigma} \ , \nonumber \\
&& \delta \; C^2_{\mu\nu,\rho} \ = \Box \Lambda^2_{\mu\nu,\sigma} \ , \nonumber \\
&& \delta \; D^{11}_{\rho\sigma} \ = \ \partial^\mu\;
\Lambda^1_{\mu,\rho\sigma} \, + \, \partial_\rho\;
\Lambda^{112}_\sigma \, + \, \partial_\sigma\; \Lambda^{112}_\rho
\ , \nonumber \\
&& \delta \; D^{12}_{\mu,\rho} \ = \ \partial^\nu\;
\Lambda^2_{\mu\nu,\rho} \, - \, \partial_\mu\; \Lambda^{112}_\rho \ , \nonumber \\
&& \delta \; D^{21}_{\mu,\rho} \ = \ \partial^\sigma\;
\Lambda^1_{\mu,\rho\sigma} \, + \, \partial_\rho\; \Lambda^{122}_\mu \ , \nonumber \\
&& \delta \; D^{22}_{\mu\nu} \ = \ \partial^\rho\;
\Lambda^2_{\mu\nu,\rho} \, - \, \partial_\mu\; \Lambda^{122}_\nu
\, - \, \partial_\nu\; \Lambda^{122}_\mu \ . \label{BRSTgentrg}
\end{eqnarray}
A partial gauge fixing of eqs. (\ref{gentriplet}) and
(\ref{BRSTgentr}), making use of the new gauge parameters, reduces
the fields to the naive set, but the constraints
(\ref{gentrconst}) now indeed emerge as additional field equations
introduced by the BRST procedure.

This more general set of fields exhausts the spectrum of the open
bosonic string in the tensionless limit. Actually, the generalized
triplets (\ref{gentrip}) exhaust all cases presented by the closed
bosonic string as well, since for the closed string the relevant
states involve at least two sets of oscillators associated with
left and right world-sheet modes. Hence, the spectra of all
bosonic models in the tensionless limit are built out of an
infinite collection of these (generalized) triplets.

\section{(A)dS extension of bosonic string triplets}

In this Section we describe how to construct the (A)dS extension
of the massless triplets that have emerged from the bosonic string
in the tensionless limit, but for brevity we confine our attention
to symmetric triplets. It is well known that higher-spin gauge
fields propagate consistently and independently of one another in
conformally flat space times, thus bypassing the well-known
Aragone-Deser inconsistencies \cite{ades} that would be introduced
by a background Weyl tensor. The bosonic triplets also allow this
extension rather simply, but it is instructive to see how the
story develops.

\subsection{Direct construction}

One can construct directly the (A)dS extensions of the bosonic
triplets, starting from the flat-space equations
(\ref{flattriplet}) and (\ref{flattripletgauge}). While the gauge
transformations of $\varphi$ and $D$ are naturally turned into
their curved-space counterparts
\begin{eqnarray}
&& \delta \varphi \ = \ \nabla \Lambda \ , \nonumber \\
&& \delta D \ = \ \nabla \cdot \Lambda \ , \label{deltaphidads}
\label{adstripletgauge}
\end{eqnarray}
the key observation is to deduce the deformed transformation of
$C$ from the condition that the constraint relating it to
$\varphi$ and $D$, \be C \ = \ \nabla \cdot \varphi \ - \ \nabla D
\ , \ee be retained.

The result,
\begin{equation}
\delta C
 \ = \ \Box \ \Lambda \ + \ \frac{(s-1)(3-s-{\cal D})}{L^2}\
  \Lambda \ +
\ \frac{2}{L^2} \,  g \; \Lambda^{'} \ , \label{deltacads}
\end{equation}
where ``primes'' as usual denote traces, ${\cal D}$ is the
space-time dimension, $L^2$ determines the (A)dS cosmological
constant, $g$ denotes the background metric tensor and $s$ denotes
the spin of $\varphi$, then fixes unambiguously the form of the
other equations. All this rests on the only new datum of the
deformed problem, the commutator of two covariant derivatives on a
vector,
 \be  [ \nabla_\mu ,
\nabla_\nu ] \, V_\rho \ = \ \frac{1}{L^2} \left( g_{\nu\rho} \,
V_\mu \ - \
 g_{\mu\rho} \, V_\nu \right) \ . \label{noncommder} \ee
This result actually applies to AdS, while the corresponding one
for a dS background can be formally recovered continuing $L^2$ to
negative values.

The gauge transformations (\ref{adstripletgauge}) and
(\ref{deltacads}) determine completely the resulting (A)dS
equations, that can be presented in the rather compact form
\begin{eqnarray}
&& \Box \; \varphi \ = \ \nabla C + \frac{1}{L^2} \, \left\{ 8 \, g
\, D \ - \ 2 \, g \, \varphi^{'} \ + \ \left[ (2-s)(3-{\cal D}-s) -s \right]
\,
\varphi \right\} \nonumber \ , \\
&& C = \nabla \cdot \varphi - \nabla D \nonumber \ , \\
&& \Box \; D \ = \ \nabla \cdot C \ + \frac{1}{L^2} \left\{
[s({\cal D}+s-2) +6] D - 4 \varphi^{'} - 2 g D^{'} \right\} \ .
\label{AdStriplet}
\end{eqnarray}

As in the previous section, one can also eliminate $C$. To this
end, it is convenient to define the $AdS$ Fronsdal operator
\begin{equation}
{\cal F} \ = \ \Box \; \varphi \ - \ \nabla \nabla \cdot \varphi \ + \
\frac{1}{2} \, \{ \nabla, \nabla \} \, \varphi^{'} \ ,
\end{equation}
and the first equation then becomes
\begin{eqnarray}
{\cal F} &=& \frac{1}{2} \, \{ \nabla, \nabla \} \, \left(
\varphi^{'} - 2 D \right)+ \frac{1}{L^2} \, \biggl\{ 8 \, g \, D \ -
\
2 \, g \, \varphi^{'}  \nonumber \\
&+&  \ \left[ (2-s)(3-{\cal D}-s) -s \right] \, \varphi \biggr\} \ .
\end{eqnarray}
In a similar fashion, after eliminating the auxiliary field $C$
the (A)dS equation for $D$ becomes
\begin{eqnarray}
\Box \; D \ + \ \frac{1}{2} \; \nabla \nabla \cdot D &-&
\frac{1}{2} \; \nabla \cdot \nabla \cdot \varphi \ = \ - \
\frac{(s-2)(4-{\cal D}-s)}{2L^2} \ D \ - \ \frac{1}{L^2} \ g \ D^{'}
\nonumber \\
&+& \ \frac{1}{2 L^2} \; \left\{ [s({\cal D}+s-2) +6] D - 4 \varphi^{'} - 2
g D^{'} \right\} \ .
\end{eqnarray}

It is also convenient to elaborate further on these expressions,
defining the modified Fronsdal operator
\begin{equation}
{\cal F}_L \ = \ {\cal F} \ - \ \frac{1}{L^2} \, \left\{ \left[
(3-{\cal D}-s)(2-s) - s \right]\, \varphi \ + \ 2 \, g \;
\varphi^{'} \right\} \ ,
\end{equation}
since in terms of ${\cal F}_L$ the deformed Bianchi identity
(\ref{bianchiflat}) retains a rather simple form,
\begin{equation}
\nabla \cdot {\cal F}_L \ - \ \frac{1}{2} \, \nabla \, {\cal
F}_L^{'} \ = \ - \ \frac{3}{2} \ \nabla^3 \ \varphi^{''} \ + \
\frac{2}{L^2} \ g \ \nabla\, \varphi^{''} \ . \label{bianchiads}
\end{equation}
In Section 4 we shall see how this Bianchi identity determines
\emph{local} non-Lagrangian higher-spin equations in (A)dS with
the same unconstrained gauge symmetry present in the non-local
geometric construction of \cite{fs1,fs2}.

\subsection{Consistency and the AdS BRST charge}

We have thus seen how the triplets emerging from the bosonic
string in the tensionless limit extend rather simply to the case
of (A)dS backgrounds, although the tensile string spectrum does
not display such a simple behavior for well-known reasons related
to the central extension of the Virasoro algebra. It is very
instructive to retrace these steps in the BRST formulation, since
the resulting analysis clarifies the reasons behind the very
consistency of the construction. Indeed, as we shall see in later
sections, the fermionic triplets proposed in \cite{fs2} can be
derived from the tensionless limit of the fermionic string, but do
not allow a similar Lagrangian (A)dS deformation for reasons that
the BRST analysis will explain rather neatly.

The starting point for this discussion is the (A)dS form of the
commutator of two covariant derivatives on a vector of eq.
(\ref{noncommder}). In trying to adapt the BRST construction to
this case, let us begin by introducing the tangent-space valued
oscillators $(\alpha_{-1}^a,\alpha_1^a)$, that satisfy \be [
\alpha_{1}^a , \alpha_{-1}^b ] \ = \ \eta^{ab} \ , \ee or the
corresponding oscillators $(\alpha_{-1}^\mu,\alpha_1^\mu)$,
obtained contracting them with the vielbein $e_\mu^a$, that
satisfy \be [ \alpha_{1}^\mu , \alpha_{-1}^\nu ] \ = \ g^{\mu\nu}
\ , \ee where $g$ denotes the (A)dS metric.

The ordinary partial derivative must now be replaced by an
operator that, acting on the totally symmetric Fock-space tensors
built from the single oscillator $\alpha_{-1}$, produces the
proper covariant derivative. This operator, denoted in the
following again by $p_\mu$, can be defined as \be p_\mu \ = \  -\;
i \, \left(
\partial_\mu - \Gamma^\rho{}_{\mu\nu} \, \alpha_{-1}^{\nu}\,
 \alpha_{1 \; \rho} \right)\label{covflat}
 \ ,
\ee or equivalently as \be p_\mu \ = \  -\; i \, \left(
\partial_\mu + \omega_{\mu}^{ab} \, \alpha_{-1 \; a}\,
 \alpha_{1 \; b} \right) \ , \label{covcurve}
\ee where $\Gamma$ and $\omega$ denote the Christoffel and spin
connections. It is then simple to verify that \be [p_\mu,p_\nu] \
= \ \frac{1}{L^2}\; (\alpha_{-1\; \mu} \, \alpha_{1\; \nu} \, -\,
\alpha_{-1\; \nu} \, \alpha_{1\; \mu}) \ , \ee since for an (A)dS
space the Riemann tensor is simply \be R_{\mu\nu\rho\sigma} \ = \
\frac{1}{L^2} \, \left( g_{\mu\rho} \, g_{\nu\sigma} \ - \
g_{\nu\rho} \, g_{\mu\sigma} \right) \ . \ee In a similar fashion,
one can see that \be \label{lapl} \ell_0 \ = \ g^{\mu \nu} \, (
p_\mu p_\nu \, + \, i\; \Gamma^\lambda_{\mu \nu}\; p_\lambda )\ =
\ p^a \; p_a \, - \ i \, \omega_a{}^{ab} \, p_b \ee acts on
Fock-space tensors as the proper D'Alembertian operator.

 In order to determine the (A)dS extension of the
BRST charge (\ref{Qlimbose}), let us insist on retaining the two
constraints associated to\footnote{The operators $\ell_{\pm 1}$
are hermitian conjugates of one another with respect to the AdS
integration measure.} \be \label{nogh} \ell_{\pm 1} \ = \
\alpha_{\pm 1} \cdot p \ , \ee now built with the covariant
derivative operator (\ref{covflat}) or, equivalently,
(\ref{covcurve}). However, the commutator of $\ell_1$ and
$\ell_{-1}$ does not generate $\ell_0$ as in flat space. Rather,
\be [ \ell_1 , \ell_{-1} ] \ = \ \tilde{\ell_0} \ , \ee where the
modified D'Alembertian is \be \tilde \ell_0 \ = \ \ell_0 \ - \
\frac{1}{L^2}\, \left( -{\cal D} \, + \, \frac{{\cal D}^2}{4} \, +
\, 4 \, M^\dagger \; M \, - \, N^2 \, + \, 2\, N \right) \ , \ee
with ${\cal D}$, as in previous sections, the total space-time
dimension. Here \be N \ = \ \alpha_{-1} \cdot \alpha_{1} \ + \
\frac{\cal D}{2} \ \label{Nop} \ee is like the contribution of
$\alpha_{\pm 1}$ to the squared mass in the tensile $L_0$
generator, and thus counts the number of indices of the Fock-space
fields, up to the space-time dimension ${\cal D}$, while \be M \ =
\ \frac{1}{2}\; \alpha_{1} \cdot \alpha_{1} \ \label{Mop} \ee is
like the contribution of $\alpha_{1}$ to the tensile $L_2$
generator, and thus takes traces of the Fock-space fields.

The emergence of these new operators enlarges the algebra, that
now includes the additional commutators
\begin{eqnarray} \label{al}
&&
[ M^\dagger \; , \; \ell_1] \ = \ -\, \ell_{-1} \ , \nonumber \\
&&[\tilde \ell_0 \; , \; \ell_1] \ = \ \frac{2}{L^2}\, \ell_1 \, -
\, \frac{4}{L^2}N \, \ell_1 \, + \,
      \frac{8}{L^2}\, \ell_{-1} \; M \ , \nonumber\\
&&[ N \; , \; \ell_1 ] \ = \ -\; \ell_1  \ , \label{tripletalg1}
\end{eqnarray}
and their hermitian conjugates, together with
\begin{eqnarray} \label{so21}
&&[N \; , \; M] \ = \ - \; 2\; M \ , \nonumber \\
&&[M^\dagger \; , \; N] \ = \ - \, 2\; M^\dagger \ , \nonumber
\\
&&[M^\dagger \; , \; M] \ = \ -\, N \ , \label{tripletalg2}
\end{eqnarray}
that define an $SO(1,2)$ subalgebra.

Notice that (\ref{tripletalg1}) and (\ref{tripletalg2}) is
actually a \emph{non-linear algebra}, and therefore the associated
BRST charge should be naively constructed with the recipe of
\cite{jtm}. As in \cite{bpt}, however, this would introduce a
larger set of ghosts and corresponding fields, going beyond the
triplet structure. Thus, in the spirit of the flat limit for the
triplet, \emph{let us retain only the $(\ell_{\pm 1},\ell_0)$
constraints, treating (\ref{tripletalg1}) as an ordinary algebra
where $M$, $M^\dagger$ and $N$ play the role of ``structure
constants''}. Remarkably, this is possible and guarantees the
Lagrangian nature of eqs. (\ref{AdStriplet}) for the deformed
triplets, since the additional operators act ``diagonally'' on the
triplet fields, their only effect being to mix them and to
introduce in the resulting equations some coefficients that depend
explicitly on the spin $s$ and on the space-time dimension ${\cal
D}$.

With this proviso, one can write the \emph{identically nilpotent}
BRST charge
\begin{eqnarray} \label{brst} \nonumber
Q& =&c_0\; \left(\tilde{\ell}_0 \, - \, \frac{4}{L^2}  N \, + \,
\frac{6}{L^2} \right)
  \, + \, c_1\; \ell_{-1}\  \, + \, c_{-1} \; \ell_1
    \, - \, c_{-1} \; c_1 \; b_0 \\ \nonumber
   &-& \, \frac{6}{L^2} \; c_0\; c_{-1} \; b_1
      \, - \, \frac{6}{L^2} \; c_0 \; b_{-1} \; c_1
        \, + \, \frac{4}{L^2} \; c_0 \; c_{-1} \; b_1 \; N
      \, + \, \frac{4}{L^2} \;c _0 \; b_{-1} \;  c_1 \; N \\
&-& \,\frac{8}{L^2} \; c_0 \; c_{-1} \; b_{-1} \; M \, + \,
\frac{8}{L^2} \; c_0 \; c_1 \; b_1 \; M^\dagger \, + \,
\frac{12}{L^2} \; c_0 \; c_{-1} \; b_{-1} \; c_1 \; b_1 \ .
\end{eqnarray}
The nilpotency of $Q$ ensures the consistency of the construction,
and as usual determines a BRST invariant Lagrangian of the form
(\ref{LBRST}), and thus a Lagrangian set of equations as in
(\ref{stringeq}). In component notation
\begin{eqnarray} \nonumber
{\cal L} &=& - \, \frac{1}{2}\ (\nabla_\mu \varphi)^2 \ + \ s\,
\nabla \cdot \varphi \, C \ + \ s(s-1)\, \nabla \cdot C \, D \
 + \ \frac{s(s-1)}{2} \, (\nabla_\mu D)^2 \ - \ \frac{s}{2} \,
C^2 \\ \nonumber &+& \ \frac{s(s-1)}{2L^2}\, {(\varphi^{'})}^2 \ -
\ \frac{s(s-1)(s-2)(s-3)}{2L^2} \, {(D^{'})}^2
  \ - \ \frac{4s(s-1)}{L^2} \, D \, \varphi^{'} \\
&-& \ \frac{1}{2L^2} \, \left[ (s-2)({\cal D}+s-3) \, - \, s \right]
{\varphi}^2 \ + \ \frac{s(s-1)}{2L^2} \, \left[ s({\cal D}+s-2)+6
\right]\,  D^2 \ , \label{LtripletBADS}
\end{eqnarray}
whose field equations are indeed those in (\ref{AdStriplet}).

\section{Compensator form of the  bosonic higher-spin equations}

In this Section we show how one can obtain local non-Lagrangian
descriptions of higher-spin bosons that exhibit the unconstrained
gauge symmetry present in the non-local geometric equations of
\cite{fs1,fs2} and reduce to the Fronsdal form after a partial
gauge fixing. The triplets are actually very useful in this
respect, since they suggest directly the form of the resulting
equations. One can also arrive at more complicated fully gauge
invariant Lagrangian formulations for higher-spin bosons, that are
nicely determined by an extension of the BRST method discussed in
the previous sections, obtained enlarging the constraint algebra
as in \cite{pt2}. Whereas the resulting equations were there
connected to the Fronsdal formulation, here we shall see that a
suitable partial gauge fixing and the corresponding judicious
elimination of a number of auxiliary fields recovers the
unconstrained gauge symmetry of \cite{fs1,fs2}, and thus the
non-Lagrangian equations presented in the next subsection.

\subsection{Non-Lagrangian formulation in flat space}

The case of a single propagating spin-$s$ field can be recovered
from the results of the previous section restricting the attention
to field configurations such that all lower-spin excitations are
pure gauge. To this end, it suffices to demand that
\begin{equation}
\varphi^{'} \ - \ 2\, D \ = \ \partial \, \alpha \ ,
\label{maxspin}
\end{equation}
where $\alpha$ is a spin-$(s-3)$ field that will play the role of
the single compensator needed in this formulation. This choice,
motivated by the fact that $\varphi^{'} - 2 D$ transforms as a
canonical spin-$(s-2)$ field, turns the first of eqs.
(\ref{tripletnoB}) into \be {\cal F} \ = \ 3 \,
\partial^{\; 3} \, \alpha \ , \label{flatlocal}
\ee while the second eq. (\ref{tripletnoB}) takes an apparently
more complicated form, and becomes
\begin{equation}
\Box \, \varphi^{'} \ + \ \frac{1}{2}\, \partial \ \partial \cdot
\varphi^{'} \ - \ \partial \cdot \partial \cdot \varphi \ = \
\frac{3}{2} \, \Box \, \partial \alpha \ + \ \partial^2 \
\partial \cdot \alpha \ . \label{deq}
\end{equation}
In terms of the Fronsdal operator defined in eq.
(\ref{fronsdalflat}), however, this simplifies considerably, since
(\ref{fronsdalflat}) implies that
\begin{equation}
{\cal F}^{'} \ = \ 2 \, \Box \, \varphi^{'} \ - \ 2\, \partial
\cdot
\partial \cdot \varphi \ + \ \partial \ \partial \cdot \varphi^{'} \ + \ \partial^2 \
\varphi^{''}\ ,
\end{equation}
so that eq. (\ref{deq}) is equivalent to
\begin{equation}
{\cal F}^{'} \ - \ \partial^2 \, \varphi^{''} \ = \ 3\, \Box
\partial \alpha \ + \ 2 \, \partial^2 \ \partial \cdot \alpha \ .
\end{equation}
On the other hand, the trace of eq. (\ref{flatlocal}) is
\begin{equation}
{\cal F}^{'} \ = \ 3\, \Box \partial \alpha \ + \ 6 \,
\partial^2 \ \partial \cdot \alpha \ + \ 3\,
\partial^3 \alpha^{'} \ ,
\end{equation}
and thus, by comparison, one obtains
\begin{equation}
\partial^2 \, \varphi^{''} \ = \ 4 \, \partial^2 \ \partial \cdot \alpha \
+ \ \partial^3 \ \alpha^{'} \ = \ \partial^2\ \left( 4
\partial \cdot \alpha \ + \ \partial \alpha^{'} \right) \ .
\end{equation}

The conclusion is that the triplet equations imply a pair of
\emph{local} equations for a single massless spin-$s$ gauge field
$\varphi$ and a single spin-$(s-3)$ compensator $\alpha$,
\begin{eqnarray}
&& {\cal F} \ = \ 3\, \partial^{\; 3} \, \alpha \ , \nonumber \\
&& \varphi^{''} \ = \ 4\,  \partial \cdot \alpha \ + \ \partial \,
\alpha^{'} \ , \label{flatbosecompens}
\end{eqnarray}
that are invariant under the \emph{unconstrained} gauge
transformations
\begin{eqnarray}
&& \delta \; \varphi \ = \ \partial \; \Lambda \ , \\
&& \delta \; \alpha \ = \ \Lambda^{'} \ ,
\end{eqnarray}
and clearly reduce to the standard Fronsdal form after a partial
gauge fixing using the trace $\Lambda^{'}$ of the gauge parameter.
These equations are nicely consistent, since the second is
implied by the first, as can be seen using the Bianchi identity of
eq. (\ref{bianchiflat}). However, \emph{these are not Lagrangian
equations}, somewhat in the spirit of the Vasiliev form of
higher-spin dynamics \cite{vas,ss}.

\subsection{Non-Lagrangian formulation in (A)dS}

One can also obtain the (A)dS extension of the spin-$s$
compensator equations (\ref{flatbosecompens}). To this end, the
starting point are the (A)dS gauge transformations for the fields
$\varphi$ and $\alpha$, that in such a curved background take
naturally the form
\begin{eqnarray}
&& \delta \; \varphi \ = \ \nabla \Lambda \ , \nonumber \\
&& \delta \; \alpha \ = \ \Lambda^{'} \ .
\end{eqnarray}
One can then proceed in various ways, for instance starting from
the gauge variation of the (A)dS form of the Fronsdal operator
\begin{eqnarray}
\delta {\cal F}_L &\equiv& \delta \; \left\{ {\cal F} -
\frac{1}{L^2}\,
[(3-{\cal D}-s)(2-s) -s] \varphi - 2 g \varphi^{'} \right\} \nonumber \\
&=& 3 \; ( \nabla^3 \Lambda^{'}) \, - \, \frac{4}{L^2}\ g \nabla
\Lambda^{'} \ ,
\end{eqnarray}
and it is then simple to conclude that the compensator form of the
higher-spin equations in (A)dS is
\begin{eqnarray}
&& {\cal F} \ = \ 3 \nabla^3 \alpha + \frac{1}{L^2}\ \{ [(3-{\cal
D}-s)(2-s) -s] \varphi + 2 g \varphi^{'} \} - \frac{4}{L^2}\ g
\nabla
\alpha \ , \nonumber \\
&& \varphi^{''} \ = \ 4 \nabla \cdot \alpha \ + \ \nabla
\alpha^{'} \ . \label{AdScompens}
\end{eqnarray}
These are again nicely consistent: making use of the Bianchi
identity of eq. (\ref{bianchiads}) one can in fact verify that the
first of (\ref{AdScompens}) implies the second. However,
Lagrangian equations can be obtained, both in flat space and in an
(A)dS background, from a BRST construction based on a wider set of
constraints, an issue to which we now turn.

\subsection{BRST analysis and compensator Lagrangian in flat space}

The previous constructions show that the BRST machinery encodes
quite neatly the physical state conditions one wants to describe
for these systems, and on the other hand provides a direct path
toward their inclusion in suitable off-shell formulations. Thus
for the flat-space triplet one builds the BRST operator, as in
\cite{pt1}, out of the three generators $(\ell_0,\ell_{\pm 1})$,
and this leads to a description where the field $\varphi$ is
eventually subject to the conditions \be \Box \; \varphi \ = \ 0 \
, \qquad \qquad \
\partial \cdot \varphi \ = \ 0 \ ,
\label{stphysical} \ee that indeed propagate a chain of modes of
spins $s$, $s-2$, \ldots, $0$ or $1$ according to whether $s$ is
even or odd. In a similar fashion, the description of irreducible
spin-$s$ modes would require the additional on-shell constraint
\begin{equation}
\varphi^{'} \ = \ 0 \ , \end{equation} and this would bring about
the operators $M$ and $M^\dagger$ that we have already met in
subsection 3.2. However, while there we treated them as structure
constants of the $(\ell_0,\ell_{\pm 1})$ triplet algebra, here we
shall introduce corresponding ghost-antighost pairs $(c_{\pm M} ,
b_{\pm M} )$ , that as usual satisfy the anti-commutation
relations \be \{ \; c_{\pm 1\, M} \; , \; b_{\mp 1\, M}\; \} \ = \
1 \ , \ee and resort to the construction of \cite{pt2}, whose
result is indeed an off-shell system that embodies the compensator
equations (\ref{flatbosecompens}). A more complicated BRST
construction, described in \cite{bpt}, adapted to the non-linear
constraint algebra (\ref{tripletalg1}) and (\ref{tripletalg2}),
would also determine the AdS deformation of this system, that we
shall not discuss for brevity.

Let us therefore begin by reviewing the results in \cite{pt2},
whose BRST procedure rests on the algebra
\begin{eqnarray}
&&[\ell_{1}\; , \;M^\dagger] \ = \ \ell_{-1}\ , \qquad
[\ell_{-1}\; , \;M] \ = \ - \ \ell_{1} \ , \qquad
[\ell_1\;,\;\ell_{-1}] \ = \ \ell_0 \ , \nonumber \\
&&[M\; , \;M^\dagger] \ = \ N \ , \qquad [M,N]\ =\ 2\, M \ ,
\qquad [M^\dagger,N]\ =\ -\, 2 \, M^\dagger \ ,\label{algecomp}
\end{eqnarray}
where $N$ and $M$ are defined in eqs. (\ref{Nop}) and (\ref{Mop}).
Notice that the new operators, $M$, $M^\dagger$ and $N$, close on
an SO(1,2) subalgebra. The BRST construction for this system
presents an interesting subtlety, since $N$, a \emph{strictly
positive} operator, cannot be regarded as providing a physical
state condition in the spirit of (\ref{stphysical}). Hence,
although the algebra is formally closed, it effectively includes
second-class constraints associated with $M$ and $M^\dagger$,
whose commutator gives rise to the offending operator $N$. A way
out, however, is provided in \cite{pt2,bpt}, whose basic idea is
to eliminate the offending constraint via an auxiliary realization
of the algebra involving an additional oscillator,
 $d$, that we shall
take to satisfy the commutation relation \be [d,d^\dagger]=-1 \ ,
\label{dosc} \ee and an additional parameter, $h$, that plays the
role of the non-trivial dynamical value of the offending
constraint. In practice, one can dispose of this constraint
altogether, rotating it away by a suitable unitary transformation
built from the conjugate momentum of $h$. The states in the
enlarged Fock space are expanded, as usual, in (anti)ghost modes,
and each of the resulting terms,
\begin{equation}
|\varphi_{i}\rangle \ = \ \sum_k \; |\varphi_{i,k}\rangle \ \equiv
\ \sum_k \; \varphi^{k}_{i;\mu_1 \mu_2...\mu_p}\;
\alpha_{-1}{}^{\mu_1} \; \alpha_{-1}{}^{\mu_2} \; ...\;
\alpha_{-1}{} ^{\mu_p} \; (d^{\; \dagger})^{k} \; |0\rangle \ ,
\end{equation}
comprises arbitrary powers of the new $d^{\; \dagger}$ oscillator.
Although this expansion is formally an infinite series, as we
shall see the number of powers of $d^{\; \dagger}$ needed to
describe a spin-$s$ field is actually finite.

More specifically, in order to eliminate the offending constraint
$N$, one first modifies it, including in it a parameter $h$, to be
regarded as an additional phase-space variable. If this were done
naively, however, the algebra \p{algecomp} would not be preserved.
The way out is precisely to introduce the additional degrees of
freedom associated to the oscillator $d$ and to build an auxiliary
realization for the algebra. Clearly this complication is not
needed for $\ell_0$ and $\ell_{\pm 1}$, that are first-class
constraints. On the other hand, $M^\pm$ and $N$ close on an
SO(2,1) subalgebra, for which a convenient recipe is available,
precisely in terms of the single new oscillator $d$ of eq.
(\ref{dosc}). It is in fact simple to verify that
\begin{eqnarray}
&& M_{(aux)} \ = \ d\, \sqrt{h + 1+ d^\dagger d }  \ ,
\nonumber \\
&& M^\dagger_{(aux)} \ = \ d^\dagger\, \sqrt{h + d^\dagger d }  \
, \nonumber \\
&& N_{(aux)} \ = \ - \, 2 \, d^\dagger d \ - \ h
\end{eqnarray}
close on the SO(2,1) algebra (\ref{algecomp}). Since they clearly
commute with the original $M$ and $N$ operators, one can define
new operators, \begin{equation} \tilde{M}_\pm \ = \ M^\pm \ + \
M^\pm_{(aux)} \ , \qquad \tilde N \ = \ N \ + \ N_{(aux)} \ ,
\end{equation}
that realize again the SO(2,1) algebra  (\ref{algecomp}). The
nilpotent BRST charge for the resulting system is then formally
constructed, treating all operators under consideration as first
class constraints, as
\begin{eqnarray}
Q&=&c_0 \ell_0 \, + \, c_1 \; \ell_{-1} \, + \, c_M \; \tilde
M^\dagger \, + \, c_{-1} \; \ell_1 \, + \, c_{-M} \; \tilde M +\,
c_N \tilde N
 \nonumber \\
&-&  \, c_{-1}\; c_1 \; b_0 \, + \, c_{-1} \; b_{-1}\; c_M \, -\,
  c_{-M} \; c_1 \; b_1 \nonumber \\
&+& c_{N} (2c_{-N}b_{N} +  2b_{-N}c_{N} + c_{-1}b_{1} +
b_{-1}c_{1} -3) - c_{-M}c_{M}b_{N} \ .
\end{eqnarray}
The final step is the elimination of the term proportional to
$c_{N}$ while maintaining the nilpotency of the BRST charge. This
can be done performing on the BRST charge the unitary
transformation
\begin{equation}
Q \ \rightarrow \ e^{-i \; \pi \; x_h} \, Q \, e^{i \; \pi \; x_h}
\ , \label{unitary}
\end{equation}
where $x_h$ is the phase-space coordinate conjugate to $h$, so
that \be [x_h, h] \ = \ i \ , \ee and
\begin{equation}
\pi \ = \ M \ - \ 2 \, d^\dagger d \ + \ 2\, c_{-N}\, b_{N} \ +\
2\, b_{-N}\, c_{N} \ + \ c_{-1}\, b_{1} \ + \ b_{-1}\, c_{1} \ -\
3 \
\end{equation}
is essentially a number operator. Notice that this transformation
removes all terms depending on $c_{N}$ from the BRST charge, while
obviously preserving its nilpotency. Finally, the term containing
$b_{N}$ can be also dropped without any effect on the nilpotency,
and one is left with a BRST charge without the offending
constraint, but where the other constraints are suitably redefined
by (\ref{unitary}).

Therefore, after the unitary transformation that rotates away the
offending constraint, the \emph{identically nilpotent} BRST charge
for this system takes the form
\begin{equation}
Q \ = \ Q_1 \ + \ Q_2 \ ,
\end{equation}
with
\begin{equation}
\{ \; Q_1 \; , \; Q_2 \; \} \ = \ 0 \ , \qquad  \quad Q_1^2 \ = \
- \; Q_2^2 \ ,
\end{equation}
where
\begin{eqnarray}
Q_1&=&c_0 \ell_0 \, + \, c_1 \; \ell_{-1} \, + \, c_M \; M^\dagger
\, + \, c_{-1} \; \ell_1 \, + \, c_{-M} \; M
 \nonumber \\
&-&  \, c_{-1}\; c_1 \; b_0 \, + \, c_{-1} \; b_{-1}\; c_M \, -\,
  c_{-M} \; c_1 \; b_1 \ ,
\end{eqnarray}
and
\begin{eqnarray}
\!\!\!Q_2&=&c_{-M} \; \sqrt{\; -\; 1\; +\; N\; -\; d^\dagger \, d
\; + \; 2\; b_{-M}\; c_M \; + \; 2\; c_{-M}\; b_{M} \; + \;
b_{-1}\; c_1 \; + \; c_{-1}\; b_1 } \ \, d
\nonumber \\
&&\!\!\!\!\!\!\!\!\!\!\!\!\!\!\! +\, d^\dagger \sqrt{\; -\; 1 \; +
\; N\; - \; d^\dagger \; d \; + \; 2\; b_{-M}\; c_M \; + \; 2\;
c_{-M}\; b_{M} \; + \; b_{-1}\; c_1 \; + \; c_{-1}\; b_1 }\ \, c_M
\, .
\end{eqnarray}
Again, this determines a BRST invariant Lagrangian of the type
(\ref{LBRST}), and now the most general expansions of the state
vector $|\Phi\rangle$ and of the gauge parameter $|\Lambda
\rangle$ in ghost variables are
\begin{eqnarray}
|\Phi\rangle&=&|\varphi_1\rangle \, + \, c_{-1} \; b_{-1} \;
|\varphi_2\rangle \, + \,  c_{-M} \; b_{-M} \; |\varphi_3\rangle
\, + \, c_{-1} \; b_{-M} \; |\varphi_4\rangle \nonumber
 \\
&+& \, c_{-M} \; b_{-1} \; |\varphi_5\rangle \, + \, c_{-1} \;
c_{-M}\; b_{-1} \; b_{-M} \; |\varphi_6\rangle \, +\,
c_0 \; b_{-1} \; |C_1\rangle \nonumber\\
&+& c_0 \; b_{-M} \; |C_2\rangle \, + \, c_0 \; c_{-1} \; b_{-1}
\; b_{-M} \; |C_3\rangle \, + \, c_0 \; c_{-M} \; b_{-1} \;
b_{-M}\; |C_4\rangle \ ,
\end{eqnarray}
and
\begin{eqnarray}
|\Lambda\rangle&=&b_{-1} \, |\Lambda_1\rangle \, + \, b_{-M} \;
|\Lambda_2 \rangle \, + \,  c_{-1} \; b_{-1} \; b_{-M} \,
|\Lambda_3\rangle \, + \, c_{-M} \; b_{-1} \; b_{-M} \,
|\Lambda_4\rangle \nonumber
 \\
&+& \, c_{0} \; b_{-1} \; b_{-M} \, |\Lambda_5\rangle  \ ,
\end{eqnarray} where $|\varphi_i\rangle$ and
$|C_i\rangle$ have ghost number $g=-1/2$ and depend \emph{only} on
the bosonic creation operators $\alpha_{-1}^\mu$ and $d^\dagger$.
Let us also note that both the Lagrangian and the gauge
transformations are not affected by redefinitions of the gauge
parameters of the type \be \delta \; |\Lambda\rangle \ =\ Q \;
|\omega\rangle \ , \ee and in particular with \be |\omega\rangle\
= \ b_{-1} \; b_{-M} \; |\omega_1\rangle \ . \ee As a result, one
of the gauge parameters, $|\Lambda_5\rangle$, is inessential and
can be ignored.

With this proviso, the resulting Lagrangian in the bosonic
Fock-space notation is
\begin{eqnarray}
{\cal L} &=& \, - \, \langle C_1|C_1\rangle \,- \, \langle
C_2|\varphi_2\rangle \, + \, \langle C_3|\varphi_3\rangle\, + \,
\langle C_4|C_4\rangle\, - \,
  \langle \varphi_2|C_2\rangle \, + \, \langle \varphi_3|C_3\rangle  \nonumber \\
&-& \langle C_1| M^\dagger|\varphi_4\rangle \, - \, \langle C_1|
\ell_{-1}|\varphi_2\rangle \, + \,
  \langle C_1| \ell_1|\varphi_1\rangle \, - \, \langle C_2| M^\dagger|\varphi_3\rangle \,
  - \, \langle C_2| \ell_{-1} |\varphi_5\rangle  \nonumber \\
\nonumber &+&\langle C_2| M|\varphi_1\rangle \, - \, \langle C_3|
M^\dagger|\varphi_6\rangle \, + \, \langle C_3|
\ell_1|\varphi_5\rangle \, - \, \langle C_3| M|\varphi_2\rangle \,
+\, \langle C_4| \ell_{-1}|\varphi_6\rangle \\
\nonumber &+&\langle C_4| \ell_1|\varphi_3\rangle \, - \, \langle
C_4| M|\varphi_4\rangle \, + \, \langle \varphi_1|
M^\dagger|C_2\rangle \, + \, \langle \varphi_1|
\ell_{-1}|C_1\rangle \, -\,
\langle \varphi_1| \ell_0|\varphi_1\rangle \\
\nonumber &-&\langle \varphi_2| M^\dagger|C_3\rangle \, + \,
\langle \varphi_2| \ell_0|\varphi_2\rangle \, - \, \langle
\varphi_2| \ell_1|C_1\rangle \, + \, \langle \varphi_3|
\ell_{-1}|C_4\rangle \, + \,
\langle \varphi_3| \ell_0|\varphi_3\rangle   \\
\nonumber &-&\langle \varphi_3| M|C_2\rangle \, - \, \langle
\varphi_4| M^\dagger|C_4\rangle \, + \, \langle \varphi_4|
\ell_0|\varphi_5\rangle \, - \, \langle \varphi_4| M|C_1\rangle \,
-\,
\langle \varphi_5| \ell_{-1}|C_3\rangle  \\
\nonumber &+&\langle \varphi_5| \ell_0|\varphi_4\rangle \, - \,
\langle \varphi_5| \ell_1|C_2\rangle \,- \, \langle \varphi_6|
\ell_0|\varphi_6\rangle \, + \, \langle \varphi_6|
\ell_1|C_4\rangle \, - \,
\langle \varphi_6| M|C_3\rangle  \\
\nonumber &-&\langle C_1| d^\dagger X_{1}|\varphi_4\rangle \, - \,
\langle C_2| d^\dagger X_{2}|\varphi_3\rangle \, + \, \langle C_2|
X_{0}\; d|\varphi_1 \rangle
\, - \, \langle C_3| d^\dagger X_{1,4}|\varphi_6\rangle \\
\nonumber &-&  \langle C_3| X_{2}\; d|\varphi_2\rangle \, -\,
\langle C_4| X_{3}\; d|\varphi_4\rangle \, + \, \langle \varphi_1|
d^\dagger X_{0}|C_2\rangle \, - \, \langle \varphi_2| d^\dagger X_{2}|C_3\rangle \\
\label{CL} &-&\langle \varphi_3| X_{2}\; d|C_2\rangle \, - \,
\langle \varphi_4| d^\dagger X_{3}|C_4\rangle \, - \, \langle
\varphi_4| X_{1}\; d|C_1\rangle\, - \, \langle \varphi_6| X_{4}\;
d|C_3\rangle \ ,
\end{eqnarray}
where
\begin{equation}
X_{r}\ = \ \sqrt{-1 + N  - d^\dagger d + r} \ .
\end{equation}
In the compact index-free tensorial notation, the same Lagrangian
reads
\begin{eqnarray}
{\cal L} &=& \sum_k \ \Biggl[ Y_{k,0}\, \varphi_1^k \, \Box \,
\varphi_1^k \, - \, Y_{k,2}\, \varphi_2^k \, \Box \, \varphi_2^k
\, - \, Y_{k,4}\, \varphi_3^k \, \Box \, \varphi_3^k \, -\,
Y_{k,3}\, \varphi_4^k \, \Box \, \varphi_5^k\, -\, Y_{k,3}\,
\varphi_5^k \, \Box \, \varphi_4^k \nonumber \\&+& Y_{k,6}\,
\varphi_6^k \, \Box \, \varphi_6^k \, - \, Y_{k,1} \, {(C_1^k)}^2
 \, + \, 2 \, Y_{k,1} \, C_2^k \, \varphi_2^k \, - \,
2\, Y_{k,4} \, C_3^k \, \varphi_3^k \, + \, Y_{k,5} \, {( C_4)}^2
\nonumber \\ &-&  Y_{k,3} \, (C_1^k)' \, \varphi_4^k \, - \, 2\,
Y_{k,1} \, C_1^k \,
\partial \, \varphi_2^k \, - \, 2\, Y_{k,0} \, C_1^k\,
\partial \, \varphi_1^k \, + \, Y_{k,4} \, (C_2^k)' \, \varphi_3^k\nonumber \\
 &-& 2\, Y_{k,2} \, C_2^k \, \partial \, \varphi_5^k \, - \, Y_{k,2} \, C_2^k \, (\varphi_1^k)'
 \, + \, Y_{k,6}\, ( C_3^k)'
\, \varphi_6^k \, - \, 2\, Y_{k,3}\, \varphi_5^k \, \partial C_3^k
\nonumber \\ &+& Y_{k,4}\, C_3^k \, (\varphi_2^k)' \, + \, 2 \,
Y_{k,5} \, C_4^k \,
\partial \,
\varphi_6^k \, - \, Y_{k,4} \, \varphi_3^k \, \partial C_4^k \, -
\, 2 \, Y_{k,5} \, C_4^k \, (\varphi_4^k)' \nonumber \\ &-& 2\,
\sqrt{s-k-3+\frac{\cal D}{2}}\ \left( - Y_{k,3} C_1^{k+1} \,
\varphi_4^k \, + \, Y_{k,4} \, C_2^{k+1} \, \varphi_3^k \, -
\, Y_{k,2} \, C_2^k \, \varphi_1^{k+1}\right. \nonumber \\
&+& \left. Y_{k,6}\, C_3^{k+1}\, \varphi_6^k \, + \, Y_{k,4} \,
C_3^k \, \varphi_2^{k+1} \, + \, Y_{k,5}\, C_4^k \,
\varphi_4^{k+1}\right)\Biggr] \ ,
\end{eqnarray}
with
\begin{equation}
Y_{k,r} \ = \ \frac{{(-1)}^k}{(s-2k-r)!} \ .
\end{equation}

The complete field equations are then
\begin{eqnarray}                     \label{equation2}
&&   -\, \eta \, C_2^k \ - \ \partial C_1^k \ + \ \Box\,
\varphi_1^k
\ - \ k\, \sqrt{s-k-2 +\frac{\cal D}{2}} \ C_2^{k-1} \ = \ 0 \ , \nonumber \\
&&   C_2^k \ + \ \eta \, C_3^k \ - \ \Box \, \varphi_2^k
\ +\ \partial \cdot C_1^k \ + \ k\,
\sqrt{s-k -2 +\frac{\cal D}{2}} \ C_3^{k-1} \ = \ 0 \ ,\nonumber\\
&&   - \, C_3^k \ - \ \partial \, C_4^k \ - \ \Box\,
\varphi_3^k \ + \ \frac{1}{2}(C_2^k)'
 \ - \ \sqrt{s-k-3 \ + \ \frac{\cal D}{2}} \ C_2^{k+1} \ = \ 0 \ ,\nonumber\\
&& \eta \, C_4^k \, + \, \Box \, \varphi_5^k \, + \, \frac{1}{2}\,
(C_1^k)' \, + \, k \, \sqrt{s-k-2+\frac{\cal D}{2}} \,
C_4^{k-1}\, - \, \sqrt{s-k-3+\frac{\cal D}{2}} \, C_1^{k+1} \, = \, 0 \ ,\nonumber\\
&& \partial \, C_3^k \ + \ \Box \, \varphi_4^k  \ - \
\partial \, \cdot C_2^k \ = \ 0 \ ,
\nonumber\\
&&   \Box \, \varphi_6^k \ - \ \partial \cdot C_4^k \ +
\ \frac{1}{2}\, (C_3^{k+1})'
\ -\ \sqrt{s-k-3 +\frac{\cal D}{2}} \ C_3^{k+1} \ = \ 0 \ ,\nonumber\\
&& C_1^k
 \ + \ \eta \varphi_4^k  \ + \ \partial \varphi_2^k
\ - \ \partial \cdot \varphi_1^k \ - \ k\, \sqrt{s-k-2+\frac{\cal D}{2}} \ \varphi_4^{k-1} \ = \ 0 \ ,\nonumber\\
&&    \eta \, \varphi_3^k
 \, - \, \partial \, \varphi_5^k \, - \, \frac{1}{2}\, (\varphi_1^k)'
\, + \, k \, \sqrt{s-k-2 +\frac{\cal D}{2}} \, \varphi_3^{k-1}
 \, - \, \sqrt{s-k-3 +\frac{\cal D}{2}} \, \varphi_1^{k+1} \, +  \, \varphi_2^k \, = \, 0 \ ,
\nonumber\\
&&    \eta \, \varphi_6^k \, + \, \partial \cdot \varphi_5^k
 \, + \, \frac{1}{2}\, (\varphi_2^k)'
\, + \, k \, \sqrt{s-k-2 +\frac{\cal D}{2}} \, \varphi_6^{k-1}
 \, - \, \sqrt{s-k-3 + \frac{\cal D}{2}} \, \varphi_2^{k+1} \, - \, \varphi_3^k \, = \, 0 \ ,
\nonumber\\
&& C_4^k  \ - \ \partial \, \varphi_6^k  \ + \ \partial \cdot
\varphi_3^k \ - \ \frac{1}{2}\, (\varphi_4^k)' \ + \
\sqrt{s-k-3+\frac{\cal D}{2}}\ \varphi_4^{k+1} \ = \ 0 \ ,
\end{eqnarray}
where $\eta$ denotes the Minkowski metric, while the corresponding
gauge transformations are
\begin{eqnarray}\label{gauge2}
\delta \, \varphi_1^k& =&
\partial \, \Lambda_1^k \ +\
\eta \, \Lambda_2^k \ +\
k\, \sqrt{s-k -2 +\frac{\cal D}{2}} \ \Lambda_2^{k-1}\ ,\\
\delta \, \varphi_2^k &=&
 \Lambda_2^k \ +\
\partial \cdot \Lambda_1^k \ +
\ \eta \, \Lambda_3^k \ + \ k\,
\sqrt{s-k -2 +\frac{\cal D}{2}} \ \Lambda_3^{k-1} \ , \nonumber\\
\delta \, \varphi_3^k &=& -\Lambda_3^k \ + \ \frac{1}{2}
(\Lambda_2^k)' \ -\
\partial \, \Lambda_4^k \ -\
\sqrt{s-k-3 +\frac{\cal D}{2}}\
 \Lambda_2^{k+1}  \ ,\nonumber\\
\delta \, \varphi_4^k &=&  \partial \cdot \Lambda_2^k\
- \ \partial \, \Lambda_3^k \ ,\nonumber\\
 \delta \varphi_5^k &=& - \, \eta\,
  \Lambda_4^k \ - \ \frac{1}{2}\,
(\Lambda_1^k)' \ -\ k \, \sqrt{s-k-2 +\frac{\cal D}{2}} \
\Lambda_4^{k-1}\
+ \ \sqrt{s-k-3 +\frac{\cal D}{2}} \ \Lambda_1^{k+1} \  , \nonumber\\
\delta \, \varphi_6 &=& - \, \frac{1}{2}(\Lambda_3^k)' \ + \
\partial \cdot \Lambda_4^k
\ + \ \sqrt{s-k-3 +\frac{\cal D}{2}} \ \Lambda_3^{k+1} \ , \nonumber\\
\delta \, C_1^k &=&
 \Box \, \Lambda_1^k \ , \nonumber\\
\delta \, C_2^k &=& \Box \, \Lambda_2^k  \ , \nonumber\\
\delta \, C_3^k &=&
\Box \, \Lambda_3^k   \ , \nonumber\\
\label{gauge10} \delta \, C_4^k &=&
 \Box \, \Lambda_4^k \ .
\end{eqnarray}

From the field equations and the gauge transformations one can
unambiguously read the oscillator content of the vectors
$|\varphi_i \rangle$, $|C_i \rangle$ and $| \Lambda_i \rangle$. In
order to describe a spin-$s$ field, let us fix the number of
oscillators $\alpha_{-1}^\mu$ in the zeroth-order term of the
expansion of $| \varphi_1 \rangle$ in the oscillator $d^\dagger$,
that we shall denote by $\varphi_1^0$, to be equal to $s$. This is
actually the field $\varphi$ of the previous subsections, while
all other terms describe auxiliary or compensator fields. The
zeroth-order components in the $d^\dagger$ oscillators for the
other fields have thus the following $\alpha_{-1}^\mu$ content,
here summarized in terms of the resulting total spin, displayed
within brackets: $\varphi_2^0 \; [s-2]$ , $\varphi_3^0\; [s-4]$ ,
$\varphi_4^0\; [s-3]$ ,
 $\varphi_5^0\; [s-3]$ ,  $\varphi_6^0\; [s-6]$ ,  $C_1^0\; [s-1]$ ,
 $C_2^0\; [s-2]$ ,
 $C_3^0\; [s-4]$ ,  $C_4^0\; [s-5]$ ,  $\Lambda_1^0\; [s-1]$ ,
 $\Lambda_2^0\; [s-2]$ ,
 $\Lambda_3^0\; [s-4]$ ,  $\Lambda_4^0\; [s-5]$.
Moreover, the field equations and the gauge transformations show
that each power of the $d^\dagger$ oscillator reduces the number
of $\alpha_{-1}^\mu$ oscillators by two units, so that, for
instance, the $\varphi_1^k$ component field has $s-2k$ oscillators
of this type, and thus spin $(s-2k)$. Therefore, as anticipated,
in this off-shell formulation a spin-$s$ field requires finitely
many auxiliary fields and gauge transformation parameters,
although their total number grows linearly with $s$.

Combining the gauge transformations with the field equations, it
is possible to choose a gauge where all fields aside from
$\varphi_1^0, \varphi_2^0, \varphi_5^0$ and $C_1^0$ are
eliminated, so that one is left with a reduced set of equations
invariant under an unconstrained gauge symmetry of parameter
$\Lambda_1^0$. To this end, one first gauges away all fields
$C_i^k$ but $C_1^0$, and the residual gauge transformations are
restricted by the conditions \be \ell_0 \, \Lambda_1^k \ = \ 0
\quad (k \neq 0)\quad  {\rm and}\quad  \ell_0 \Lambda_i^k \ = \ 0
\quad (i=2,3,4) \quad {\rm and} \quad k \geq 0 \ . \ee The
parameters $\Lambda_1^k$ $(k \neq 0)$ and $\Lambda_4^k$ gauge away
$\varphi_5^k$ $(k \neq 0)$, while the parameters $\Lambda_3^k$
gauge away $\varphi_6^k$. The conclusion is that one is finally
left with gauge transformation parameters restricted by the
additional condition \be( M \ + \ X_{4} \, d ) \, |\Lambda_3
\rangle \ = \ 0 \ , \ee and with the help of these parameters
$|\Lambda_2 \rangle$ and $|\Lambda_3 \rangle$ one can also gauge
away
 $\varphi_1^k$, $\varphi_2^k$ $(k \neq 0)$ and $\varphi_3^k$, while
$\varphi_4^k$ vanishes as a result of the field equations.

One can now identify $\varphi_1^0$ with $\varphi$, $\varphi_2^0$
with $D$, $C_1^0$ with $C$, $-2\; \varphi_5^0$ with the
compensator $\alpha$ and $\Lambda_1^0$ with the gauge parameter of
the previous subsections. The first, second, seventh and eighth
equations in (\ref{equation2}) then produce the triplet and
compensator equations of the previous subsections, while the
fourth and ninth equations are consequences of these.

This construction is clearly somewhat complicated with respect to
the non-Lagrangian equations (\ref{flatbosecompens}). For
instance, the off-shell description of a spin-4 field
$\varphi_{\mu\nu\rho\sigma}$ makes use of thirteen different
fields, out of which, however, eight are of $D$ type and five are
of $C$ type, that can be simply eliminated. One can use the
additional gauge parameters and field equations to eliminate all
fields aside from the original $\varphi$, its two triplet partners
and the compensator, whose relation to the triplet fields is now
one of the residual equations of motion rather than a constraint
as in subsection 4.1. For brevity, we refrain from discussing the
(A)dS extension of these results, that is similarly related to the
analysis in \cite{bpt}. The far simpler triplets, as we have seen,
provide an alternative description of irreducible higher-spin
multiplets in (A)dS backgrounds.

\section{The fermionic triplets}

We can now turn to the fermionic triplets, that were proposed in
\cite{fs2} as a natural guess for the field equations of symmetric
spinor-tensors arising in the tensionless limit of superstrings.
As we shall see, they indeed emerge in this limit, although, as is
usually the case for the fermionic string, the dominant types of
fields are (generalized) forms rather than symmetric tensors.
Actually, we shall not be able to pursue the analysis to the same
level of detail as in the previous sections. Thus, while we shall
derive both triplet and Vasiliev-like compensator equations for
higher-spin fermionic gauge fields, we shall not be able to
present a corresponding compensator Lagrangian formulation, since
it is being constructed by other authors using the same BRST
approach discussed in the previous section \cite{bpnew}. Moreover,
we shall not be able to extend the fermionic triplets to off-shell
systems in (A)dS, and here the BRST analysis will explain clearly
the difficulty, related to the nature of the algebra of the
resulting deformed constraints.

\subsection{Open superstring oscillators}

Most of the results of the previous sections can be naturally
extended to superstrings. For brevity, we restrict our attention
to the open sector of the type-I superstring, but closed
superstrings could be treated in a similar way. Let us first
perform the $\alpha^\prime \rightarrow \infty$ limit in the BRST
charge for the open superstring
\begin{eqnarray} \nonumber
Q&=& \sum_{-\infty}^{+\infty}
      \, \left[ L_{-n} \, C_n \ + \ G_{-r}\, \Gamma_r
\ - \ \frac{1}{2}\;
(m-n):C_{-m}\, C_{-n}\, B_{m+n}: \right. \\
&+& \left. \left(\frac{3 n}{2} \ + \ m \right) :C_{-n}\, {\cal
B}_{-m}\, \Gamma_{m+n}: \ - \ \Gamma_{-n}\, \Gamma_{-m} \, {\cal
B}_{m+n} \right] \ - \ a \, C_0 \ ,
\end{eqnarray}
where $a$ is the intercept and the super-Virasoro generators
\begin{eqnarray} \label{VS}
&& L_k \ = \ \frac{1}{2}\, \sum_{l= -\infty}^{+\infty}\,
\alpha_{k-l}\, \alpha_l \ + \ \frac{1}{4}\, \sum_r \, (2r-l)\,
\psi_{l-r}\, \psi_r\ ,\nonumber
\\
\label{VF} && G_r \ = \ \sum_{l=-\infty}^{+\infty} \alpha_l \,
\psi_{r-l}\ ,
\end{eqnarray}
obey the super-Virasoro algebra
\begin{eqnarray}
&&[L_k, L_l] \ = \ (k-l)\, L_{k+l} \ +  \ \frac{{\cal D}}{8}\, \,
(k^3-k) \ ,
\nonumber \\
&&[L_k, G_r] \ = \ \left( \frac{k}{2} \, -\, r \right)\, G_{k+r} \
,
\nonumber \\
&&  \{ G_r, G_s\} \ =\ 2\, L_{r+s} \ +\ \frac{{\cal D}}{2}\,
\left(r^2 \, - \, \frac{1}{4}\right) \, \delta_{rs} \ .
\end{eqnarray}
Here $(k,l)$ are integers for both the Neveu-Schwarz (NS) and
Ramond (R) sectors, while $(r,s)$ are integers for the R sector
and half-odd integers for the NS sector, ${\cal D}$ denotes once
more the space-time dimension (${\cal D}=10$ for the tensile
string) and
 $\alpha_0^\mu = \sqrt{2 \alpha^\prime} \, p^\mu $.
The fermionic oscillators $\psi^\mu_r$ and the ghosts $\Gamma_r$
and antighosts ${\cal B}_r$ satisfy
\begin{equation}
\{ \psi^\mu_r, \psi^\nu_s \} \ = \ \delta_{r+s,0}\, \eta^{\mu \nu} \ ,
\qquad
[\Gamma_r , {\cal B}_s] \ = \ i \, \delta_{r+s,0}\ ,
\end{equation}
and the intercept is $ a=0$ in the R
sector and  $ a=\frac{1}{2}$ in the NS sector.

Rescaling the ghost variables as \be \gamma_{-r} \ = \
\sqrt{2\alpha^\prime} \ \Gamma_{-r}\ , \qquad \beta_{r} \ =\
\frac{1}{\sqrt{2\alpha^\prime}}\ {\cal B}_r \ee and then taking
the $\alpha^\prime \rightarrow \infty$ limit, one obtains the
nilpotent BRST charge for the NS sector \be \label{BRSTNS} Q_{NS}
\ = \ c_0 \, \ell_0 \ + \ \tilde Q_{NS} \ - \ M_{NS} \, b_0 \ ,
\ee with
\begin{eqnarray}
&& \label{TQ} \tilde Q_{NS}\ = \ \sum_{k \neq 0} \; \left [\;
c_{-k} \, \ell_k  \, + \, \gamma_{-r}\, g_r \right ]\ , \nonumber \\
\label{TN} && M_{NS}\ = \ \frac{1}{2} \, \sum_{-\infty}^{+\infty}
\; \left [ \, k \, c_{-k}\, c_k \, +\, \gamma_{-r} \, \gamma_r \,
\right ] \ ,
\end{eqnarray}
and
\be \label{GB}
 g_r \ = \ p \cdot \psi_{r} \ . \ee
In a similar fashion, the limiting BRST charge for the R sector
reads
 \be Q_R \ = \ c_0 \, \ell_0 \ + \ \gamma_0 \, g_0 \ + \ \tilde Q_R \ - \ M_R
b_0 \ - \ \frac{1}{2}\, \gamma_0^2 \, b_0 \ , \ee where $\tilde
Q_R$ and $M_R$ are again given by \p{TN},
 the only difference being that their sums are over
half-odd integer modes for fermionic Virasoro generators and
bosonic (anti)ghosts. Both BRST charges are again identically
nilpotent, independently of the space-time dimension ${\cal D}$.

For the type I superstring,
the string field is invariant under the action
of the BRST invariant GSO projection operators
 \be \label{GSONS}
 P_{NS} \ = \
\frac{1}{2} \, \left[ 1 \ - \ (-1)^{\psi_p^\dagger \; \psi_p \,
+\, i \gamma_p^\dagger \; \beta_p \, - \, i\; \gamma_p\;
\beta_p^\dagger}\right] \ee and
 \be \label{GSOR}
 P_R \ = \ \frac{1}{2}\,\left[ 1\ +\
\gamma_{11}\, (-1)^{\psi^\dagger_r \; \psi_r \, + \, i\;
\gamma^\dagger_r \; \beta_r \, - \, i \; \gamma_r \;
\beta^\dagger_r \, + \, i \; \gamma_0 \; \beta_0}\right] \ , \ee
where $\gamma_{11}$ is the ten-dimensional chirality matrix, that
apply to the NS and R sectors respectively. Expanding the NS
string field and the gauge parameter in terms of the fermionic
ghost zero mode as
\begin{eqnarray}
&& |\Phi^{NS} \rangle \ = \ |\Phi_1^{NS}\rangle \ + \ c_0|\Phi_2^{NS}
\rangle \ , \nonumber \\
&& |\Lambda^{NS} \rangle \ = \ |\Lambda_1^{NS} \rangle
\ + \ c_0 |\Lambda_2^{NS} \rangle  \ , \end{eqnarray}
and making use of the BRST charge
\p{BRSTNS}, one obtains the field equations
\begin{eqnarray} \label{EMNS1}
&& \ell_0 |\Phi_1^{NS} \rangle \ - \ \tilde Q_{NS}|\Phi_2^{NS} \rangle \ = \ 0
\ , \nonumber  \\
&& \tilde Q_{NS} |\Phi_1^{NS} \rangle \ - \ M_{NS} |\Phi_2^{NS}
\rangle \ =\ 0 \ , \label{EMNS2} \end{eqnarray} along with the
gauge transformations
\begin{eqnarray}
&& \delta |\Phi_1^{NS} \rangle \ = \ \tilde Q_{NS}
|\Lambda_1^{NS} \rangle \ - \ M_{NS} |\Lambda_2^{NS} \rangle \ , \nonumber \\
\label{NSG3}
&& \delta |\Phi_2^{NS} \rangle \ =\
\ell_0 |\Lambda_1^{NS} \rangle \ - \ \tilde Q_{NS}|\Lambda_2^{NS} \rangle \ .
\end{eqnarray}
The R sector is more complicated, due to the presence of the
bosonic ghost zero mode $\gamma_0$. However, one can work with the
truncated string field
\begin{equation} \label{truncated}
|\Phi^R \rangle \ = \ |\Phi_1^R\rangle \ + \ \gamma_0 \,  |\Phi_2^R \rangle
\ + \ 2 \, c_0 \, g_0\, | \Phi^R_2\rangle \ ,
\end{equation}
while still preserving the relevant portion of the gauge symmetry
and, of course, not affecting the physical spectrum \cite{KNNW1}.
The resulting, consistently truncated, field equations
\begin{eqnarray} \label{EMR1}
&& g_0 \, |\Phi_1^R \rangle \ + \ \tilde Q_{R}|\Phi_2^R \rangle \ =\ 0 \ ,
\nonumber \\
\label{EMR2} && \tilde Q_{R} \, |\Phi_1^R \rangle \ - \ 2 \, M_R
\, g_0 \, |\Phi_2^R \rangle \ = \ 0 \ ,
\end{eqnarray}
are then invariant under the gauge transformations
\begin{eqnarray} \label{GTR1}
&& \delta\, |\Phi_1^R\rangle \ = \ \tilde Q_R |\Lambda_1^R \rangle
\ +\ 2 \,M_R \, g_0 \, |\Lambda_2^R \rangle \ , \label{GT1}
\nonumber \\
\label{GTR2}
&& \delta|\Phi_2^R\rangle \ = \ g_0 \, |\Lambda_1^R \rangle \ - \ \tilde
Q_R \, |\Lambda_2^R \rangle \ . \label{GT2}
\end{eqnarray}

\subsection{Symmetric spinor-tensors}
If, as for the bosonic string, one considers fields $|\Phi^{R,1}
\rangle$ and  $|\Phi^{R,2} \rangle$ depending only on the bosonic
oscillator $\alpha^\mu_{-1}$ and on the fermionic ghost variables
$c_{-1}$ and $b_{-1}$, the expansions
\begin{eqnarray}
&& |\Phi_1^R \rangle \ = \ \frac{1}{n!}\ \psi_{\mu_1\mu_2\, ...\,
\mu_n}(x)\, \alpha^{\mu_1}_{-1} \, \alpha^{\mu}_{-1} \, ...\,
\alpha^{\mu_n}_{-1}\, |0\rangle \nonumber \\ \nonumber && \qquad
\qquad +\ \frac{1}{(n-2)!}\ \lambda_{\mu_1\mu_2\, ...\,
\mu_{n-2}}(x)\, \alpha^{\mu_1}_{-1} \, \alpha^{ \mu_2}_{-1} \,
..\, \alpha^{ \mu_{n-2}}_{-1} \, |0\rangle \ ,
\\
&&|\Phi_2^R\rangle \ = \ - \ \frac{1}{\sqrt{2}\; (n-1)!}\
\chi_{\mu_1\mu_2\, ...\, \mu_{n-1}}(x) \, \alpha^{\mu_1}_{-1} \,
\alpha^{ \mu_2}_{-1} \, ...\, \alpha^{\mu_{n-1}}_{-1}\, |0\rangle
\end{eqnarray}
define spinor-tensor fields $\psi$, $\chi$ and $\lambda$ totally
symmetric in their tensor indices and of spin $(n+1/2)$, $(n-1/2)$
and $(n-3/2)$, respectively. Substituting these expressions in the
field equations \p{EMR1} - \p{EMR2} then yields precisely the
fermionic triplet equations of \cite{fs2}:
\begin{eqnarray}
&& \dsll \psi \ = \ \partial \chi \ , \nonumber \\
&& \partial \cdot \psi \ - \ \partial \lambda
\ = \ \dsll \chi \ , \nonumber \\
&& \dsll \lambda \ =\
\partial \cdot \chi \ . \label{fermitriplet}
\end{eqnarray}
The BRST gauge invariance involves an unconstrained parameter,
\begin{equation}
|\Lambda^\prime_1 \rangle \ =\ \frac{1}{(n-1)!}\
\epsilon_{\mu_1\mu_2...\mu_{n-1}}(x) \, \alpha_{-1}^{\mu_1} \,
\alpha_{-1}^{\mu_2} ... \alpha_{-1}^{\mu_{n-1}}\, |0\rangle \ ,
\end{equation}
and determines the gauge transformations
\begin{eqnarray}
&& \delta \psi \ = \ \partial \, \epsilon \ , \nonumber \\
&& \delta \Lambda \ = \ \partial \cdot \epsilon \ ,
\nonumber \\
&& \delta  \chi \ = \ \dsll \epsilon \ ,
\end{eqnarray}
in agreement with \cite{fs2}.

Let us note, however, that the totally symmetric bosonic triplets
of subsection 2.3  do not arise directly in the NS sector of the
open superstring, since all states containing only bosonic
$\alpha^\mu$ oscillators and fermionic $b,c$ ghosts are eliminated
by the GSO projection operator \p{GSONS}. However, they can emerge
from tensors with mixed symmetry, or even directly if the GSO
projection is modified to correspond to type-0 strings
\cite{type0}. Generalized triplets of mixed symmetry are actually
the superpartners of symmetric fermionic triplets in the type-I
superstring.

One can also consider generalized triplets for spinor-tensors,
that also arise in the R sector, and these, described by
\begin{eqnarray}
&& g_0\; |\lambda_{i_1...i_l}^{j_1..j_l} \rangle \ - \ {(-1)}^l\,
\ell_{i_l}\; |\chi_{i_1...i_{l-1}}^{j_1..j_l} \rangle \ + \
{(-1)}^l\,
\ell_{-j}\; |\chi_{i_1...i_l}^{j j_1..j_l} \rangle \ = \ 0 \ , \nonumber \\
&& \ell_{i_l}\; |\lambda_{i_1...i_{l-1}}^{j_1,..j_{l-1}} \rangle \
- \ \ell_{-j}\; |\lambda_{i_1...i_{l}}^{j j_1,..j_{l-1}} \rangle \
- \ 2\, g_0\; {(-1)}^l\; |\chi_{i_1,...i_{l-1}}^{i_l j_1
.j_{l-1}} \rangle \ = \ 0 \ ,
\end{eqnarray}
resemble the generalized bosonic triplets of subsection 2.4.

These equations follow from the Lagrangians
\begin{eqnarray}
{\cal L} &=& \sum_l \ \left[ \frac{(-1)^l}{{(l!)}^2} \langle
\lambda_{j_1...j_l}^{i_1...i_l}|\; g_0\;
|\lambda_{i_1...i_l}^{j_1...j_l} \rangle \ - \ \frac{2}{(l-1)!l!}
\; \langle \chi_{i_1...i_{l-1}}^{j_1...j_l}| \;
\ell_{-i_l} \; | \lambda_{j_1...j_l}^{i_1...i_l} \rangle  \right. \nonumber \\
&-& \left. \frac{ 2\; {(-1)^l}}{{(l!)}^2} \langle
\chi_{i_1...i_{l}}^{j_i...j_{l+1}}|\; \ell_{j_{l+1}} \;
|\lambda_{j_1...j_l}^{i_1...i_l} \rangle \ + \
\frac{{(-1)}^l}{{((l-1)!)}^2} \langle
\chi_{j_1...j_{l-1}}^{i_1...i_{l-1}j_l}\; |\;g_0\; |
 \chi_{i_1...i_{l-1}}^{j_1...j_l} \rangle \right] \ .
\end{eqnarray}
that are invariant under the gauge transformations
\begin{eqnarray}
\delta \; |\lambda_{i_1,...i_l}^{j_i,..j_l} \rangle  &=& - \,
{(-1)}^l\, \ell_{i_l}\, |\Lambda^{1(1)\;
j_1,..j_{l}}_{i_1,...i_{l-1}} \rangle \, +\, {(-1)}^l
 \ell_{-j}\; |\Lambda^{1(1)\; j,j_1,..j_{l}}_{i_1,...i_{l}}
\rangle \nonumber \\ &&+ \,2 \, g_0 \, |\Lambda^{2(1)\;
i_l,j_1,..j_{l}}_{i_1,...i_{l-1}}
\rangle \ , \nonumber \\
 \delta \; |\chi_{i_1...i_{l-1}}^{j_i...j_l} \rangle &=& g_0\;
|\Lambda^{1(1)\; j_i,..j_l}_{i_1,...i_{l-1}} \rangle \, -\,
{(-1)}^l \ell_{i_{l-1}}\; |\Lambda^{2(1)\;
j_1,..j_l}_{i_1,...i_{l-2}} \rangle \nonumber \\ &&+ \, {(-1)}^l\,
\ell_{-j}\; |\Lambda^{(1)\; j,j_1,..j_l}_{i_1,...i_{l-1}} \rangle
\ ,
\end{eqnarray}
that also allow the ``gauge-for-gauge'' transformations
\begin{eqnarray}
 \delta \; |\Lambda^{1(k)\; j_1,..j_{l+k}}_{i_1,...i_l} \rangle &=& -\,
{(-1)}^l\ell_{i_l}\; |\Lambda^{1(k+1)\;
j_1,..j_{l+k}}_{i_1,...i_{l-1}} \rangle  \ +\ {(-1)}^l
 \ell_{-j}\; |\Lambda^{1(k+1)\; j,j_1,..j_{l+k}}_{i_1,...i_{l}}
\rangle \nonumber \\ &&+ \ 2 \, |\Lambda^{2(k+11)\;
i_l,j_1,..j_{l+k}}_{i_1,...i_{l-1}}
\rangle \ , \nonumber \\
\delta \; |\Lambda^{2(k)\; j_1...j_{l+k}}_{i_1...i_{l-1}} \rangle
&=& g_0 \; |\Lambda^{1(k+1)\; j_1,..j_{l+k}}_{i_1,...i_{l-1}}
\rangle \nonumber \ - \ {(-1)}^l \ell_{i_{l-1}}\;
|\Lambda^{2(k+1)\; j_1,..j_{l+k}}_{i_1,...i_{l-2}} \rangle
\nonumber \\ &&+ \ {(-1)}^l \ell_{-j}\; |\Lambda^{2(k+1)\;
j,j_1,..j_{l+k}}_{i_1,...i_{l-1}}  \rangle \ ,
\end{eqnarray}
and so on.

The ``mixed symmetry'' of these fields is of general type, and
allowing for the possible dependence of the string field on
$\psi_{-r}$ and $\gamma_{-r}, \beta_{-r}$ in the R sector would
lead to more complicated equations with similar properties.

\subsection{Compensator form of the fermionic higher-spin equations}
Here the story parallels the discussion in subsection 4.1, since the
 fermionic Fang-Fronsdal operator \cite{fangfronsd}
\begin{equation}
{\cal S} \ = \ i \, \left( \dsll \psi \ - \ \partial \psisl
\right)
\end{equation}
varies into a term proportional to the gamma-trace of the gauge
parameter, \be \delta {\cal S} \ = \ - 2 \, i \ \partial^2 \esl \
, \ee under the gauge transformation
\begin{equation}
\delta \psi \ = \ \partial \; \epsilon \ .
\end{equation}
In addition, ${\cal S}$ satisfies the Bianchi identity \be \prd
{\cal S} \ - \ \frac{1}{2} \, \partial \ {\cal S}{\; '} \ - \
\frac{1}{2} {\not {\! \pr}} \ssl \
 = \ i \ \pr^{\; 2} \psisl\;' \ ,  \label{bianchifermi}
\ee and as a result the gauge parameter and the gauge field were
constrained in \cite{fangfronsd} to satisfy the conditions
\begin{equation}
\esl \ = \ 0 \ , \qquad \ \psisl\;' \ = \ 0 \ .
\end{equation}

As for integer-spin fields, one can eliminate these constraints
either passing to the non-local equations of \cite{fs1} or,
alternatively, introducing a single compensator field $\xi$. The
resulting equations,
\begin{eqnarray}
&& {\cal S} \ = \ - \ 2 \, i \, \partial^2 \, \xi \ , \nonumber \\
&& \psisl^{\ '} \ = \ 2 \, \partial \cdot \xi \ + \ \partial \,
\xi^{\ '} \ + \ \dsll \xisl \ , \label{compfermiflat}
\end{eqnarray}
are then invariant under the gauge transformations
\begin{eqnarray}
&& \delta \psi \ = \ \partial \, \epsilon \ , \nonumber \\
&& \delta \xi \ = \ \esl \ ,
\end{eqnarray}
involving an unconstrained gauge parameter, and are consistent,
since the first implies the second via the Bianchi identity
(\ref{bianchifermi}).

These compensator equations generalize nicely to an (A)dS
background. The gauge transformation for a spin-$s$ fermion
becomes in this case
\begin{equation}
\delta \psi \ = \ \nabla \, \epsilon \ + \ \frac{1}{2 L} \, \gamma
\; \epsilon \ ,
\end{equation}
where, as in previous sections, $\nabla$ denotes and (A)dS
covariant derivative and $L$ determines the (A)dS curvature. In
order to proceed, one needs the commutator of two covariant
derivatives on a spin-1/2 field $\eta$,
\begin{equation}
[ \nabla_\mu , \nabla_\nu ] \, \eta \ = \ - \ \frac{1}{2 L^2} \
\gamma_{\mu\nu} \ \eta \ , \label{noncommders}
\end{equation}
where $\gamma_{\mu\nu}$ is antisymmetric in $\mu$ and $\nu$ and
equals the product $\gamma_\mu \gamma_\nu$ when $\mu$ and $\nu$
are different, that can be combined with eq. (\ref{noncommder}) to
obtain the corresponding expression for fields of arbitrary
half-odd integer spins.

For a spin-$s$ fermion ($s=n+\frac{1}{2}$), where $n$ is the
number of vector indices carried by the field $\psi$, the
compensator equations in an (A)dS background are
\begin{eqnarray}
&& \left( \nablasl \, \psi \ - \ \nabla \psisl \right) \ + \
\frac{1}{2 L} \, \left[ {\cal D}\ + \ 2( n \, - \, 2) \right] \psi
\ + \ \frac{1}{2L} \, \gamma \, \psisl \nonumber \\
&& \qquad\qquad\qquad\quad = \ -\,  \{ \nabla, \nabla \} \xi \ + \
\frac{1}{L} \, \gamma \, \nabla \, \xi \ + \
\frac{3}{2 L^2} \, g \, \xi \ , \nonumber \\
&& \psisl^{\ '} \ = \ 2 \, \nabla \cdot \xi \ + \ \nablasl \xisl \
+ \ \nabla \xi^{\ '} \ + \ \frac{1}{2 L} \, \left[ {\cal D}\ + \
2( n \, - \, 2) \right] \, \xisl  \ - \ \frac{1}{2L} \, \gamma \,
\xi^{\ '} \ , \label{fermiadsnl}
\end{eqnarray}
and are invariant under
\begin{eqnarray}
&& \delta \psi \ = \ \nabla \, \epsilon \ , \nonumber \\
&& \delta \xi \ = \ \esl \ ,
\end{eqnarray}
with an unconstrained parameter $\epsilon$. Eqs.
(\ref{fermiadsnl}) are again a pair of non-Lagrangian equations,
like their flat-space counterparts (\ref{compfermiflat}), and are
again nicely consistent, on account of the (A)dS deformation of
the Bianchi identity (\ref{bianchifermi}),
\begin{eqnarray}
\nabla \cdot {\cal S} \ - \ \frac{1}{2} \, \nabla \, {\cal S}{\;
'} \ - \ \frac{1}{2} \, \nablasl  \ssl &=& \frac{i}{4L } \, \gamma
\, S^{\, \prime} \ + \ \frac{i}{4L} \left[ ({\cal D} \, - \, 2)\,
+\, 2\; (n\, - \, 1) \right] \, \ssl \nonumber \\ &+& \
 \frac{i}{2} \ \left[  \{ \nabla , \nabla \} \ - \ \frac{1}{L}\ \gamma \, \nabla
\ - \ \frac{3}{2\;L^2}\right] \, \psisl\;'
 \ ,
\end{eqnarray}
where now the Fang-Fronsdal operator ${\cal S}$ is also deformed
and becomes
\begin{equation}
{\cal S} \ = \ i\, \left( \nablasl \, \psi \ - \ \nabla \psisl
\right) \ + \ \frac{i}{2 L} \, \left[ {\cal D}\ + \ 2( n \, - \,
2) \right] \psi \ + \ \frac{i}{2L} \, \gamma \, \psisl \ .
\end{equation}

We have been unable to construct a corresponding Lagrangian AdS
deformation for generic fermionic triplets. Already for the
simplest case of a $(3/2,1/2)$ system, that involves a pair of
fields $\psi_\mu$ and $\chi$, one can write the (A)dS equations
\begin{eqnarray}
&& \nablasl \psi_\mu \ + \ \frac{{\cal D}-2}{2 L} \, \psi_\mu \ +
\ \frac{1}{2L} \, \gamma_\mu \, \psisl \ = \ \nabla_\mu \, \chi \
,
\nonumber \\
&& \nabla \cdot \psi \ + \ \frac{{\cal D}-1}{2 L} \, \psisl \ = \
\nablasl \, \chi
\end{eqnarray}
that are invariant under the gauge transformations
\begin{eqnarray}
&& \delta\; \psi_\mu \ = \ \nabla_\mu \; \epsilon \ + \
\frac{1}{2L} \, \gamma_\mu \, \epsilon \nonumber \\
&& \delta \; \chi \ = \ \nablasl \epsilon \ + \ \frac{{\cal
D}}{2L} \, \epsilon \ ,
\end{eqnarray}
but, when suitably combined, they give rise to the further
condition
\begin{equation}
\chi \ = \ \psisl \ .
\end{equation}
As for the bosonic triplet, the modes described by this system
thus reduce to a single spin multiplet, but differently from that
case the additional constraints do not arise in the gauge-fixing
procedure, but are generated by the field equations themselves.
The origin of these difficulties is clearly spelled by the BRST
analysis. In this context, a key problem one is facing is that,
after adapting the Dirac operator $g_0 = \gamma \cdot p$ to the
(A)dS background, the resulting set of operators $\ell_0$,
$\ell_{\pm 1}$ and $g_0$ does not form a closed algebra, not even
a non-linear one as was the case for integer-spin fields on AdS.
The way out would be to enlarge the constraint algebra, including
in it the additional operators $T^\pm = \gamma \cdot \alpha^\pm$
corresponding to gamma-trace conditions, as was done for the
$M^\pm$ operators corresponding to ordinary traces in subsection
3.2, and then to construct a nilpotent BRST charge at expense of
the inclusion of further ghost fields as in \cite{bpt}. While we
hope to return to this point in the near future, the additional
constraints would lead to an off-shell description of an
irreducible spin multiplet not directly related to the triplet
structures we were after in this work.

 \vskip 24pt
\begin{flushleft}
{\bf Acknowledgments}
\end{flushleft}
\vskip 12pt

We are grateful to M. Bianchi, I. Buchbinder, C.M. Hull, J.F.
Morales, A. Pashnev, P. Pasti, E. Sezgin, D. Sorokin, P. Sundell,
M. Tonin and M. Vasiliev, and especially to X. Bekaert for several
stimulating discussions. This work was supported in part by INFN,
the EC contract HPRN-CT-2000-00122, the EC contract
HPRN-CT-2000-00148, the EC contract HPRN-CT-2000-00131, the
MIUR-COFIN contract 2001-025492, the INTAS contract N 2000-254,
and the NATO contract PST.CLG.978785. A.S. and M.T. would like to
thank the Physics Departments of the University of Padova and of
the University of Rome ``Tor Vergata'' for the kind hospitality
extended to them during the course of this work.
 \vskip 24pt
\begin{flushleft}
{\bf Note added}
\end{flushleft}
\vskip 12pt

The original version of this paper contained an improper
interpretation of the consistency conditions for the gauge-fixing
of (A)dS triplets, that we interpreted as signaling that the
spectrum would collapse to irreducible spin-$s$ modes. In fact,
the problem was just the reflection of an unsuitable gauge choice.
We are very grateful to the referee for calling to our attention
the difficulties related with our interpretation.

\end{document}